\documentclass[%
 reprint,
 amsmath,amssymb,
 aps,
]{revtex4-2}

\usepackage{verbatim}
\usepackage{graphicx}
\usepackage{dcolumn}
\usepackage{bm}
\usepackage{xcolor}
\usepackage{comment}
\usepackage{siunitx}
\usepackage{multirow}
\usepackage{subfigure}
\usepackage{amsthm,amsmath,amssymb}
\usepackage{mathrsfs}
\usepackage{balance}
\usepackage{url}
\usepackage[utf8]{inputenc}

\usepackage[columnwise]{lineno}

\begin{document}

\preprint{APS/123-QED}

\title{Booster Free From Spin Resonance For Future 100~km-scale Circular e$^{+}$e$^{-}$ Colliders}


\author{Tao Chen$^{1,2}$}
\author{Zhe Duan$^1$}%
\email{duanz@ihep.ac.cn }
\author{Daheng Ji$^1$}%
\author{Dou Wang$^1$}%
\affiliation{$^1$ Key Laboratory of Particle Acceleration Physics and Technology,
Institute of High Energy Physics, Chinese Academy of Sciences,
19B Yuquan Road, Beijing 10049, China \\
$^2$ University of Chinese Academy of Sciences,
19A Yuquan Road, Beijing 10049, China
}%

\date{\today}

\begin{abstract}
Acceleration of polarized electron~(positron) beams in a booster
synchrotron may suffer from depolarization due to crossings of many
spin depolarization resonances, and this could limit its applications.
We have studied the structure of spin depolarization resonances of a
100~km-scale booster lattice of the Circular Electron Positron Collider~(CEPC). The lattice has 8 arc regions with hundreds of FODO cells, interleaved with straight sections, 
leading to a high periodicity.
Our analysis shows the contributions to the strength of intrinsic and
imperfection spin resonances add up coherently near the super-strong
resonances beyond 120 GeV, but mostly cancel out and result in generally weak resonance strengths at lower beam energies. 
Detailed simulations confirm that beam polarization can be mostly maintained in the fast acceleration to 45.6 GeV and 80 GeV, but that severe depolarization may occur at even higher energies. 
This study suggests the possibility of acceleration of polarized 
electron~(positron) 
beams to ultra-high beam energies without the help of Siberian snakes, and supports injecting highly polarized beams into the collider rings as an attractive solution for resonant depolarization measurements and longitudinally polarized colliding beam experiments for future
100~km-scale circular e$^{+}$e$^{-}$ colliders.

\end{abstract}

\maketitle


\section{INTRODUCTION}


The Circular Electron Positron Collider (CEPC)~\cite{Lou_2019,CEPCcdr,cepcsnowmass} is one of the
future e$^{+}$e$^{-}$ collider projects~\cite{Benedikt_2019,Michizono_2019,Stapnes_2019} that aim to study the properties of the Higgs boson,
a crucial cornerstone of the Standard Model. In the Conceptual Design Report~(CDR)~\cite{CEPCcdr} released in 2018, the CEPC was 
designed as a 100~km-scale double-ring collider,
to provide unprecedented
high luminosity at center-of-mass energies of 91 GeV (Z-factory), 160 GeV (W-factory), and 240 GeV (Higgs-factory via $e^{+}e^{-}\rightarrow ZH$), upgradable to 360 GeV (ttbar energy). This powerful instrument
should not only be used for precision measurements on the properties of
these elementary particles, but also for the search for new physics. Beam polarization is a key aspect of the CEPC design. On one hand, transverse beam polarization of at least 5\% to 10\% is needed to carry out resonant depolarization~(RD) measurements~\cite{vepp,lep}, which are essential for beam energy calibration at Z and W energies.
On the other hand, longitudinally polarized colliding beams could provide an extra probe for precision tests of the Standard Model and the search for new physics via colliding beam experiments. This application requires 50\% or more longitudinal polarization at the interaction points (IPs) without significantly reducing the luminosity~\cite{duan_eeFACT_2022}. A pair of 
spin rotators would be inserted around each IP to realize the longitudinal polarization. 
To this end, solenoid-based spin rotators have been successfully 
included in the CEPC CDR lattice at Z energies~\cite{xia_investigation_2022}.

There are two main approaches for preparing polarized beams to satisfy these needs.
The first is to utilize the spontaneous polarization
build-up in the collider rings due to emission of synchrotron radiation, namely the Sokolov-Ternov effect~\cite{st}. 
This approach has been studied~\cite{blondel_polarization_2019} to 
provide a few percent of vertical
beam polarization for the RD measurements at
CERN's Future Circular $e^+ e^-$ collider (FCC-ee)~\cite{FCCcdr}. 
The other approach is to generate
polarized beams from the source and inject them into the collider rings, with the potential for achieving higher levels of beam polarization without significantly sacrificing the luminosity.

The CEPC injector chain, as outlined in
the CEPC CDR~\cite{CEPCcdr},
includes
unpolarized electron and
positron sources,
a 10~GeV main linac, a full
energy booster
and transfer lines.
The booster and the collider rings
share the same tunnel
and the booster accelerates both electron and positron beams
from 10~GeV up to
the working beam energies in an alternating manner.
The feasibility of extending this design
to cover generation and transport of polarized beams throughout the
injector is being explored~\cite{duan_eeFACT_2022}.
A polarized
electron gun can be added
to produce electron beams 
with 80\% or more polarization~\cite{brachmann_polarized_2007,wang_high_2022}.
However, the development of polarized positron sources that can
simultaneously
meet the requirements of
high polarization, nC-level bunch charge and high repetition 
rate is still a technical challenge~\cite{musumeci_positron_2022}.
Nevertheless, it is still possible to generate 20\% or more
beam polarization in the positron damping ring through
the Sokolov-Ternov effect for RD measurements~\cite{duan_eeFACT_2022}.
Maintaining the beam polarization throughout the injection
chain, including the 100 km-scale booster, is essential.
Previous studies 
for the SLC~\cite{woods_polarized_1996} and ILC~\cite{moortgat-pick_polarized_2008} have shown small polarization loss in the linac
and transfer lines, but the main concern is
the potential polarization loss during acceleration in the
booster. This is the focus of this paper.

In a high-energy electron~(positron)
circular accelerator, the precession of the single-particle spin expectation value
$\vec{S}$, and of the unit-length
spin vector $\hat{S}=\vec{S}/|\vec{S}|$, of an 
electron~(positron), follows
the Thomas-BMT equation\cite{thomas}\cite{BMT},
\begin{equation}
\frac{d\hat S}{d\theta}=\left [ \vec{\Omega}_0(\theta)+ \vec{\omega}(\vec u;\theta) \right] \times \hat S. 
\label{eq:T-BMT}
\end{equation}
with 
\begin{equation}
\vec{\Omega}_0(\theta)= \vec{\Omega}_{00}(\theta)+\Delta \vec{\Omega}(\theta)
\label{eq:Yospinmotionrev}
\end{equation}
where $\vec{\Omega}_{0}(\theta)$  is the spin precession vector on the closed orbit at the azimuthal angle $\theta$.  $\vec{\Omega}_{00}(\theta)$ and $\Delta \vec{\Omega}(\theta)$ 
represent the effect
of electromagnetic fields on the design orbit,
and the impact of magnetic errors and correction fields on the spin precession vector, respectively.
The vector $\vec{\omega}(\vec u;\theta)$
describes the influence of fields
seen by particles oscillating around the closed orbit with
the 
phase space vector $\vec u$.
In the Frenet-Serret coordinate system with a right-hand basis $(\vec{e}_x,\vec{e}_y,\vec{e}_z)$,
pointing radially outwards, vertically upwards and longitudinally (clockwise), respectively,
the phase space coordinate is
expressed as $\vec u=(x, p_x, y, p_y, z, \delta)$.
$x$ and $y$ are the horizontal and vertical coordinates of the particle. The transverse phase-space momenta $P_x$ and $P_y$ are normalized by the reference momentum $P$, i.e.,
$p_x=P_x/P$ and $p_y=P_y/P$, respectively. $z=-\beta c \Delta t$ where $\Delta t$ is the time difference between the particle and the reference particle arriving at the azimuthal angle $\theta$. $\beta=v/c$, where $v$ and $c$ are the velocity of the particle and light, respectively. $\delta=\Delta{E}/{E}$ is the relative deviation of the 
from the design energy $E$.
The polarization of an electron bunch is the ensemble average of the $\hat{S}$ of the particles in the bunch.

On the closed orbit $\vec{\Omega}_{0}(\theta)$ is one-turn periodic so that there is a unit-length periodic solution of Eq.~(\ref{eq:T-BMT}), $\hat{n}_0(\theta)$ with
$\hat{n}_0(\theta+2\pi)=\hat{n}_0(\theta)$. A spin vector that is not parallel to $\hat{n}_0$ 
precesses around $\hat{n}_0$ by an angle $2\pi \nu_0$ in one revolution, where $\nu_0$ is the
closed-orbit spin tune. In a circular accelerator designed
with a planar geometry and without solenoid fields (hereafter referred as the ``planar ring''), 
$\hat{n}_0$ is close to the vertical direction, and $\nu_0 \approx G\gamma$, where $G=0.00115965219$ for electrons (positrons), and $\gamma$ is the relativistic factor
 for the design energy $E$. 

Away from the closed orbit $\hat{n}_0(\theta)$ is generalised to the special solution of Eq.~(\ref{eq:T-BMT}),
namely the field of vectors $\hat{n} (\vec{u};\theta)$ called
the invariant spin field~(ISF)~\cite{barber}, obeying the periodicity condition $\hat{n} (\vec{u};\theta + 2\pi) =\hat{n} (\vec{u};\theta)$. Assuming that the orbital motion is
integrable, then $\vec{u}$ can be expressed in terms of action-angle
variables, and 
we denote the amplitude of
the combined betatron-synchrotron motion as 
$\vec{I}=(I_{I},I_{II},I_{III})$ with $I$, $II$ and $III$ denoting the three
eigenmodes,
which reduces to
$\vec{I}=(I_x,I_y,I_z)$ in the case of weak couplings, where
$x$, $y$ and $z$ represent the horizontal, vertical and longitudinal 
dimensions, respectively.
The rate of spin precession around $\hat{n}$ is described by the 
amplitude-dependent spin tune~(or spin tune in short) $\nu_s(\vec{I})$~\cite{barber}.
The projection of the spin vector of a particle onto $\hat{n}$ is an
adiabatic invariant $J_s =\hat{S}\cdot\hat{n}$
~\cite{hoffstaetter_adiabatic_2006}.
For typical orbital amplitudes with electrons, $\nu_s(\vec{I})\approx\nu_0$.

In a planar ring, $\vec{\Omega}_{00}$ is dominated by 
the vertical guiding magnetic fields on the design trajectory. In contrast,
$\Delta \vec{\Omega}$ and $\vec{\omega}$
contain contribution from
machine imperfections and orbital oscillations, respectively.
These contributions
might perturb the particle spin motion in a resonant manner
when the spin-orbit coupling resonances~(spin resonances in short) condition is nearly satisfied,
\begin{equation}
   \nu_s =K= k + k_x\nu_x + k_y\nu_y+ k_z\nu_z
\end{equation}
where $K$ denotes the location of the resonance, $k,k_x,k_y,k_z \in \mathbb{Z}$. $\nu_x$, $\nu_y$ and $\nu_z$ are the horizontal, vertical
betatron tunes and the synchrotron tunes, respectively, which
are conventionally used in the case of weak couplings, reduced from the more general orbital tunes $\nu_{I}$, $\nu_{II}$ and $\nu_{III}$ obtained from the orbital eigen-analysis~\cite{chao_evaluation_1979}.
$\hat{n} (\vec{u};\theta)$ deviates from $\hat{n}_0(\theta)$ near these spin
resonances.
Spin resonances with $\nu_0=k, k \in \mathbb{Z}$, are called integer spin resonances. In an imperfect ring, $\vec{n}_0$ could strongly deviate from 
the vertical direction near these integer spin resonances. 
Spin resonances with $|k_x|  +|k_y| + |k_z|=1$ and $|k_x|  +|k_y| + |k_z| > 1$ are called  first-order spin resonances and 
higher-order spin resonances, respectively. 
As we elaborate later, 
in the context of polarized beam acceleration in an imperfect planar ring, 
two families of spin resonances are most important. 
The integer spin resonances $\nu_0=k$
in the tilt of $\vec{n}_0$,
mainly driven by the horizontal magnetic fields that arise from
vertical orbit offsets in quadrupoles as well as dipole roll errors,
are also conventionally called the ``imperfection resonances''~\cite{courant_acceleration_1980}~\footnote{
Note that
an integer spin resonance $\nu_0=k$ can also be a ``sideband''
spin resonance of the ``parent'' first-order spin resonance
$\nu_0\pm \nu_z=k$, or of even higher-order
``parent'' spin resonances.
}. The first-order ``parent'' spin resonances 
$\nu_0=k\pm\nu_y$, 
driven by horizontal magnetic fields that
arise from vertical betatron oscillations in quadrupoles,
are also conventionally called the ``intrinsic resonances''~\cite{courant_acceleration_1980}.

In booster synchrotrons, as the particle energy is ramped up,
so are the closed orbit spin tune
$\nu_0$ and the spin tunes
$\nu_s(\vec{I})$ of the particles. This leads to crossings of the underlying spin resonances and in the case of non-adiabatic spin resonance crossing, possible consequent decrease in the $J_s= \hat{S}\cdot\hat{n}$ of the particles
and thus depolarization. 
If the closed orbit spin tune, $\nu_0$, changes linearly
at the rate $\alpha=\frac{d\nu_0}{d\theta}\approx\frac{dG\gamma}{d\theta}$,
the depolarization in crossing a single spin resonance at
$\nu_0=K$ can be estimated with the Froissart-Stora formula~\cite{FSEquation}
\begin{equation}
    \label{eq:F-S_formula}
    \frac{P_f}{P_i}\approx 2\exp(-\frac{\pi \vert \tilde \epsilon_K\vert^2}{2\alpha})-1
\end{equation}
where $P_i$ and $P_f$ are the polarizations before and after the resonance crossing,
and $\tilde \epsilon_K$ is the spin resonance strength, which depends on
the lattice design and machine imperfections. Its detailed methods of calculation will be
described later. 
We can identify three parameter regimes in the value of $\vert \tilde \epsilon_K\vert/\sqrt{\alpha}$. Firstly, if the resonance is 
very strong, or when the acceleration is very slow, say $\vert \tilde \epsilon_K\vert/\sqrt{\alpha}>1.84$, then $|\frac{P_f}{P_i}|>99\%$ but with a change of
sign relative to the initial value, i.e., a ``spin flip'', 
and we call this the ``slow crossing'' regime.
Second, if the resonance is very weak, or when the acceleration is very
fast, say $\vert \tilde \epsilon_K\vert/\sqrt{\alpha}<0.056$, then $\frac{P_f}{P_i}>99\%$, and we call this the ``fast crossing'' regime. Third, if both conditions are not satisfied, 
$\vert \frac{P_f}{P_i} \vert$ is reduced,
and we call this the ``intermediate'' regime. 
The above analysis is for the polarization
of a single particle, but it is crucial for the
overall understanding of the maintenance of
beam polarization in high-energy booster synchrotrons. 
In these types of accelerators, there are numerous
spin resonances that must be crossed during acceleration. It is convenient 
to use Eq.~(\ref{eq:F-S_formula}) to evaluate the polarization loss due to the
crossing of each spin resonance,
if the distances between the locations of adjacent spin resonances are much larger than the strengths of these spin resonances. This method allows for
identifying the major spin resonances that may cause significant depolarization, before conducting
more detailed simulation studies. Generally speaking, depolarization during acceleration is a significant concern in the maintenance of polarization in high-energy booster synchrotrons. 

In the acceleration of polarized
proton beams in the ZGS~\cite{khoe_acceleration_1975}, the AGS~\cite{khiari_acceleration_1989} and RHIC~\cite{bai_polarized_2006}, the crossings of some spin resonances 
are in the ``intermediate'' regime where significant depolarization could occur. To mitigate this,
various techniques have been proposed and
implemented, such as tune jump~\cite{khoe_acceleration_1975}, ac dipole~\cite{RFDipole}, and the use of
Siberian snakes~\cite{derbenev_radiative_1978}. 
A Siberian snake is a device that rotates spins by $\pi$
around a specified axis in the horizontal plane while 
perturbing the orbital motion only moderately.
When introduced into a circular accelerator,
certain setups of Siberian snakes can make the closed orbit spin tune
$\nu_0$ independent of the beam energy at $0.5$, avoiding the crossing of imperfection and first-order spin resonances during acceleration. 
Partial snakes~\cite{roser_properties_1989},
which rotates spin by a fraction of $\pi$ around a horizontal axis,
can also be used to 
mitigate weak spin resonances~\cite{huang_overcoming_2007}.
Partial snakes and full snakes have been
implemented in the AGS and RHIC, respectively,
allowing for maintaining
high proton beam polarization of over 50\% up to 255~GeV~\cite{schoefer_rhic_2012}. However,
higher-order spin resonances would still be encountered in the presence of snakes,
leading to significant polarization loss
at even higher beam energies, as seen in the studies for HERA-p~\cite{hoffstaetter_high-energy_2006}.

To the best of our knowledge,
there have been only a few experiments of polarized electron
acceleration up to a few GeV beam energies, in VEPP-2M~\cite{barkov_measurement_1987}, ELSA~\cite{ELSA} and VEPP-4M~\cite{VEPP4M,zhuravlev_current_2020}, with only a few spin resonances being crossed in each case. Some of these experiments were within the ``slow crossing'' regime. Note that with the very slow resonance crossing rate $\alpha$, radiative
spin diffusion, which is absent in proton synchrotrons,
could lead to incomplete spin flip~\cite{yokoya_effects_1983}.
To avoid this, in Ref.~\cite{VEPP4M}, a partial snake using the detector solenoids was
used to preserve the beam polarization. In the
future 100~km-scale circular
e$^+$e$^-$ colliders like the CEPC and 
FCC-ee, hundreds of imperfection and intrinsic spin
resonances will be encountered during the acceleration process
in the booster. 
To combat this,
various studies~\cite{Koop_Acceleration_2016,koop_ideas_2018,nikitin_opportunities_2019,nikitin_polarization_2020}
have proposed the use of
different setups of (partial)~Siberian snakes to mitigate depolarization. However, the practical implementation of snakes
using either bending magnets or solenoids is challenging due
to the large size and cost
required for high energy electron boosters. 

In the study for the Electron Ion Collider~(EIC)~\cite{willeke_electron_2021}, 
a concept of ``spin-resonance-free electron ring injector''
was proposed~\cite{Ranjbar_2018} to avoid severe depolarization
during acceleration to the top energy of 18~GeV.
This is achieved through a clever design of the booster lattice,
that features a high effective lattice periodicity of 96, placing the super-strong spin resonances beyond 18~GeV, below which the spin resonances are generally weak. 
The result is that the crossings of these resonances are
well within the ``fast crossing'' regime, leading to minimal depolarization. This work highlights the importance of 
understanding and utilizing the structure of spin resonances. 

In this work, we investigate the depolarization effects in the CEPC booster, by using a simplified lattice and by incorporating preliminary error modeling and corrections. 
Our analysis of the structure of spin resonances reveals
similarities to previous research on the EIC booster~\cite{Ranjbar_2018},
with relatively weak spin resonances within the range
of working beam energies.
Our estimates indicate that spin resonance crossings 
mostly occur within the ``fast crossing'' regime,
resulting in minimal polarization loss during
acceleration to 45.6~GeV and 80~GeV. 
However, at higher
energies, such as during the acceleration to 120~GeV,
depolarization is more significant. These analyses are
verified by multi-particle tracking simulations
of the whole acceleration process. Our preliminary results were
presented in \cite{duan_eeFACT_2022}.

This article is structured as follows.
In Section~\ref{sec:more_theories}, we present
more detailed theories of the strength
for the imperfection resonances and the
intrinsic resonances, and analyze the structure of spin resonances
of a simple model ring for both types
of resonances.
In Section~\ref{sec:booster_lattice}
we introduce the lattice setup of the CEPC booster.
In Section~\ref{sec:spin_resonance_structure}, 
we analyze the resonance spectra
for the CEPC booster lattice, and estimate depolarization effects
for both types of resonances.
In Section~\ref{sec:simulations}, 
we launch multi-particle tracking simulations of the beam polarization transmission
in the acceleration process,
and compare the results to our analytical estimations.
The final section summarizes our findings and suggests future studies.

\section{MORE THEORIES OF THE SPIN RESONANCE STRENGTHS\label{sec:more_theories}}

Knowledge of the strength of spin resonances is essential for analyzing polarization loss using the Froissart-Stora formula.
In our analysis, we focus on planar electron rings with 
practical machine imperfections. 
We first present the analysis of the strengths of
imperfection resonances and intrinsic resonances.
In a planar ring, the spin motion
can be analyzed using a perturbative approach~\cite{yok2}
and Fourier expansion of spin perturbations~\cite{DK72},
which reveal the driving terms of these resonances.
The corresponding Fourier component of a spin resonance
can be used in place of
the resonance strength in the Froissart-Stora formula.

However, 
determining the strength
of higher-order spin resonances, which include both
higher-order Fourier components and feed-up of lower-order components~\cite{hoffstaetter_high-energy_2006},
is not straightforward.
Instead, the strength of a higher-order spin resonance
can be determined by retrieving the amplitude of the spin-tune jump through spin tracking of the process of resonance crossing~\cite{hoffstaetter_strength_2004}. 
In this paper,
we use both analytical and numerical methods to determine
the strengths of imperfection and intrinsic resonances of a lattice,
and estimate their impact on depolarization.
Additionally, we evaluate the depolarization caused by
these resonances and more general higher-order spin resonances,
using tracking simulations that take into account the full
six-dimensional orbital motion and
three-dimensional spin motion.

To understand the strengths of spin resonances,
we then follow the analysis of S. Y. Lee's for a simplified 
model ring~\cite{lee_simple_1986,leeSpinDynamicsSnakes1997},
where the strengths of imperfection and intrinsic
resonances are analytically calculated in terms of the
optics parameters of a basic cell, multiplied by
enhancement factors that arise from the summation of
periodic cells. This analysis is crucial in understanding
the location of strong spin resonances, and clarifying the
contrast between strong resonances and weak resonances.
This analysis of the structure of spin resonances will later
be applied to the CEPC booster lattice in Section~\ref{sec:spin_resonance_structure}.

\subsection{The strength of imperfection resonances}
We denote the right-handed orthonormal set of unit-length solutions to Eq.~(\ref{eq:T-BMT}) on the design orbit by $\hat{n}_{00}(\theta)$, $\hat{m}_{00}(\theta)$ and $\hat{l}_{00}(\theta)$, where $|\Delta \vec{\Omega}|=|\vec{\omega}|=0$, and define  $\hat{k}_{00}(\theta)=\hat{m}_{00}(\theta)+i\hat{l}_{00}(\theta)$.
Then
$\hat{k}_{00}(\theta)=e^{i(\Upsilon(\theta)-\Upsilon(\theta'))}\hat{k}_{00}(\theta')$,
where $\Upsilon(\theta)$ is the spin precession phase. In a planar ring, $\Upsilon(\theta)\approx\nu_0\Phi(\theta)$,
where $\Phi(\theta)=R\int _0^{\theta}\frac{1}{\rho_x}d\theta'$
is the integrated bending angle, $R$ is the average radius of
the ring, $\rho_x$ is the radius of curvature for the local orbit.
In a perfect planar ring, $\hat{n}_{00}(\theta)$ is 
vertical while $\hat{k}_{00}(\theta)$ is in the horizontal plane, respectively.
Without loss of generality, 
we choose $(\hat{n}_{00}, \hat{m}_{00}, \hat{l}_{00})$ to be aligned with 
$(-\vec{e}_y,\vec{e}_x,\vec{e}_z)$, respectively
at $\theta=0$.

In the presence of machine imperfections, we can expand $\hat{n}_0(\theta)$ as~\cite{yok2}
\begin{equation}
\hat{n}_0(\theta)\approx \hat{n}_{00}(\theta)+\mathscr{Re}(c_1(\theta) \hat{k}_{00}^{\star}(\theta))
\label{eq:n0approx}
\end{equation}
Putting Eq.~(\ref{eq:n0approx}) into the Thomas-BMT equation on the closed orbit, we find
\begin{equation}
\frac{d c_1(\theta)}{d \theta} \approx -i \Delta \vec{\Omega}(\theta) \cdot \hat{k}_{00}(\theta)
\end{equation}
so that 
\begin{equation}
c_1(\theta) \approx -i\int^{\theta}_{-\infty} e^{\epsilon \theta'} \Delta \vec{\Omega}(\theta') \cdot \hat{k}_{00}(\theta') d\theta' 
\label{eq:c1eq}
\end{equation}
where now and later, $e^{\epsilon \theta'}$ denotes an infinitesimal damping factor to make the integral well behaved as $\theta' \rightarrow -\infty$~\cite{mane_electron-spin_1987}~\footnote{A physics picture that clarifies
the integral of $\theta'$ from $-\infty$ is introduced in
~\cite{xia_evaluation_2022}
using the concept of ``anti-damping''~\cite{hoffstaetter_high-energy_2006}.}.
Note that $\vec{\Omega}(\theta) \cdot \hat{k}_{00}(\theta)e^{-i \nu_0\theta}$ is a periodic
function of $\theta$, which can be expanded into a Fourier series,
\begin{equation}
\Delta\Omega(\theta') \cdot \hat{k}_{00}(\theta')=\sum_{k=-\infty}^{\infty} {\tilde \epsilon}_k e^{i(\nu_0 - k)\theta'}
\end{equation}
with ${\tilde \epsilon}_k$ being the complex strength of the integer spin resonance $\nu_0=k$,
\begin{equation}
    {\tilde \epsilon}_k = \frac{1}{2\pi }\int^{2\pi}_{0}\Delta\Omega(\theta') \cdot \hat{k}_{00}(\theta') e^{-i(\nu_0-k)\theta'}d \theta'
\end{equation}
Then Eq.~(\ref{eq:c1eq}) can be simplified to the following form,
\begin{equation}
c_1(\theta) \approx -i\sum_{k=-\infty}^{\infty} \frac{{\tilde \epsilon}_k e^{i(\nu_0-k)\theta} }{\nu_0-k}
\label{eq:eyk}
\end{equation}
Therefore, ${\tilde \epsilon}_k$ characterizes the deviation of $\hat{n}_0$ from $\hat{n}_{00}$
near the integer spin resonances $\nu_0=k, k \in \mathbb{Z}$.

Since $\hat{k}_{00}$ is in the horizontal plane, only the component of
$\Delta \vec{\Omega}$ in the horizontal plane contributes to the ${\tilde \epsilon}_k$. In imperfect planar rings, the contribution of the radial magnetic field
dominates for most cases.
$\Delta\Omega_x$ denotes the component of $\Delta \vec{\Omega}$ along the $\vec{e}_x$ direction and represents the influence of the radial magnetic field $\Delta B_x$.
Its contribution to ${\tilde \epsilon}_k$ can be expressed as
\begin{eqnarray}
    {\tilde \epsilon}_k & \approx& \frac{1}{2\pi }\int^{2\pi}_{0}\Delta\Omega_x(\theta') e^{i(\nu_0 \Phi(\theta')-(\nu_0-k)\theta')}d \theta' \nonumber \\
    &=&-\frac{R(1+G\gamma)}{2\pi }\int^{2\pi}_{0}\frac{\Delta B_x(\theta')}{B\rho} e^{i k \Phi(\theta')}d \theta' 
\end{eqnarray}
where we use the relations $\vec{e}_x(\theta') \cdot \hat{k}_{00}(\theta')
=\vec{e}_x(0) \cdot \hat{k}_{00}(0) e^{i \nu_0 \Phi(\theta') }=e^{i \nu_0 \Phi(\theta') }$ and $\nu_0=k$. $B\rho$ is the magnetic rigidity. 

Let's denote $K=\nu_0=k$ as the location of the imperfection resonances, whose strengths ${\tilde \epsilon}_K^{\textrm{imp}}$ can be expressed as
\begin{equation}
    {\tilde \epsilon}_K^{\textrm{imp}}\approx -\frac{R(1+K)}{2\pi }\int^{2\pi}_{0}\frac{\Delta B_x(\theta')}{B\rho} e^{i K \Phi(\theta')}d \theta'
    \label{eq:imperfection_resonance_strength_0}
\end{equation}

The radial
magnetic field $\Delta B_x$ can be expanded in terms of the horizontal and vertical 
closed orbit $x_{\mathrm{co}}$
and $y_{\mathrm{co}}$ as,
\begin{equation}
    \frac{\Delta B_x}{B\rho}\approx(\frac{\Delta B_x}{B\rho})_0+
    \frac{\frac{\partial B_x}{\partial x}}{B\rho}x_{\mathrm{co}}
    +\frac{\frac{\partial B_x}{\partial y}}{B\rho}y_{\mathrm{co}}
\end{equation}
where the feed-down effects of multipoles are neglected,
and $\frac{\frac{\partial B_x}{\partial x}}{B\rho}x_{\mathrm{co}}$
and $\frac{\frac{\partial B_x}{\partial y}}{B\rho}y_{\mathrm{co}}$
describe
the influence of
horizontal closed-orbit positions inside skew quadrupoles, and
that of vertical closed-orbit positions inside normal quadrupoles, respectively.
The latter term is generally much more important.
The $(\frac{\Delta B_x}{B\rho})_0$ contains the influence
from various magnet misalignment errors, like
vertical misalignment errors of quadrupoles and roll errors of dipoles,
as well as the contribution from vertical orbital correctors. 

The magnet misalignment errors can be grouped into two different categories: random and systematic. 
Random misalignment errors typically have zero mean and their contributions to ${\tilde \epsilon}_K^{\textrm{imp}}$ tend to
cancel out for large $K$. Systematic misalignment errors, on the
other hand, arise from
variations in the vertical magnet positions after the smoothing procedure of alignment, and uneven settling of the accelerator floor over time. Their contribution to 
${\tilde \epsilon}_K^{\textrm{imp}}$ does not simply average out.
Note that these two categories of magnet misalignment errors also
have different influences on
the patterns of the orbital correctors and their contribution
to ${\tilde \epsilon}_K^{\textrm{imp}}$.
In this paper we focus on the random errors, the treatment of the systematic errors will be pursued in a separate study. Then ${\tilde \epsilon}_K^{\textrm{imp}}$ can be approximately by
\begin{equation}
    {\tilde \epsilon}_K^{\textrm{imp}}\approx -\frac{R(1+K)}{2\pi }\int^{2\pi}_{0}\frac{\frac{\partial B_x}{\partial y}}{B\rho}(\theta')y_{\mathrm{co}}(\theta') e^{i K \Phi(\theta')}d \theta'
    \label{eq:imperfection_resonance_strength}
\end{equation}
This agrees with the result obtained via the spinor algebra in Ref.~\cite{courant_acceleration_1980, leeSpinDynamicsSnakes1997}.

\subsection{The strength of intrinsic resonances}
We denote the right-handed orthonormal set of unit-length solutions to
Eq.~(\ref{eq:T-BMT}) on the closed orbit by $(\hat{n}_{0}(\theta),\hat{m}_{0}(\theta), \hat{l}_{0}(\theta))$, where
$|\Delta \vec{\Omega}|\neq 0$ while $|\vec{\omega}| = 0$, and
define $\hat{k}_0(\theta)=\hat{m}_0(\theta)+i\hat{l}_0(\theta)$.
$\hat{k}_0$ is quasi-periodic, $\hat{k}_0 (\theta+2\pi)=e^{i2\pi \nu_0} \hat{k}_0 (\theta)$.

When $|\Delta \vec{\Omega}|\neq 0$, $|\vec{\omega}|\neq 0$,
$\hat{n}(\vec u; \theta)$ can be expressed by~\cite{yok2}
\begin{equation}
\hat{n}(\vec u; \theta) =\hat{n}_0(\theta) \sqrt{1-|\zeta(\vec u; \theta)|^2}+\mathscr{Re}(\hat{k}_0^{\star}(\theta)\zeta(\vec u; \theta))
\end{equation}
where $\zeta(\vec u(\theta); \theta)$
satisfies
\begin{equation}
\frac{d\zeta}{d\theta}=-i\vec{\omega} \cdot \hat{k}_0 \sqrt{1-|\zeta|^2}+i\vec{\omega}\cdot \hat{n}_0 \zeta
\label{eq:zeta}
\end{equation}
For spin resonances up to the first-order, 
the solution for $\zeta(\vec u(\theta); \theta)$ can be expressed as 
\begin{equation}
\zeta(\vec u;\theta) \approx -i \int_{-\infty}^{\theta} 
e^{\epsilon \theta'} \vec{\omega}(\theta')\cdot \hat{k}_0(\theta') d\theta'
\label{eq:zeta0}
\end{equation}
Here $\vec{\omega}$ can be linearized with respect to the coordinates of the
betatron 
oscillations $x_{\mathrm{osc}}$ and $y_{\mathrm{osc}}$ as well as the
synchrotron oscillation $\delta_{\mathrm{osc}}$,
and decomposed into three oscillation modes:
\begin{equation}
        \vec{\omega}= \vec{\omega}_z \delta_{\mathrm{osc}}+\vec{\omega}_x x_{\mathrm{osc}}+\vec{\omega}_y y_{\mathrm{osc}}
\end{equation}
For instance, we can express $r_{\mathrm{osc}}$ with
$r=x,y$, 
in terms of the action-angle variables $I_r$ and $\psi_r$ as
\begin{equation}
    r_{\mathrm{osc}}=\sqrt{2I_r \beta_r}\cos(\psi_r+\widetilde{\Psi}_r)
\end{equation}
and where $\beta_r$ is the betatron function and $\widetilde{\Psi}_r(\theta)=\nu_r(\phi_r(\theta) - \theta)$,
with $\phi_r(\theta)=\frac{R}{\nu_r}\int^{\theta}_0 \frac{d\theta'}{\beta_r(\theta')}$.
The solution of the oscillatory motion is 
$\psi_r=\psi_{r0}+\nu_r \theta$.
Then Eq.~(\ref{eq:zeta0}) can be analyzed for each of these oscillation modes
\begin{equation}
    \zeta_r(I_r,\psi_r; \theta)= -i\sum_{\pm}\int^{\theta}_{-\infty}\vec{\omega}_r\cdot\hat{k}_0 \sqrt{\frac{I_r\beta_r}{2}}e^{\pm i(\psi_r+\widetilde{\Psi}_r)}d\theta'
\label{eq:zetaj}
\end{equation}
The integrand can be expanded into Fourier series,
\begin{equation}
    \vec{\omega}_r\cdot\hat{k}_0 \sqrt{\frac{I_r\beta_r}{2}}e^{\pm i(\psi_r+\widetilde{\Psi}_r)}=\sum_{k=-\infty}^{\infty} \tilde{ \epsilon}_{k,\pm}^{r} e^{i (\nu_0 \pm \nu_r - k)\theta'}
\end{equation}
with $\tilde{ \epsilon}_{k,\pm}^{r}$ being the complex strength of the first-order spin resonance
$\nu_0\pm\nu_r=k, k \in \mathbb{Z}$,
\begin{eqnarray}
    \tilde{ \epsilon}_{k,\pm}^{r}&=&\frac{1}{4\pi}\int_0^{2\pi}\vec{\omega}_r\cdot\hat{k}_0 \sqrt{2I_r\beta_r} \nonumber \\
    &\times& e^{ i[\pm(\psi_r+\widetilde{\Psi}_r)- (\nu_0 \pm \nu_r - k)\theta']}d\theta'
    \label{eq:first-order_resonance_strength}
\end{eqnarray}
Then Eq.~(\ref{eq:zetaj}) can be simplified to the following form,
\begin{equation}
    \zeta_r(I_r,\psi_r; \theta)= -\sum_{\pm}\sum_{k=-\infty}^{\infty} \frac{\tilde{ \epsilon}_{k,\pm}^{r} e^{i (\nu_0 \pm \nu_r - k)\theta'} }{\nu_0 \pm \nu_r - k}
    \label{eq:first-order_resonance_zeta}
\end{equation}
Therefore, $\tilde{ \epsilon}_{k,\pm}^{r}$ is a measure of the deviation of $\hat{n}$ from $\hat{n}_{0}$
near the first-order spin resonances $\nu_0\pm \nu_r=k, k \in \mathbb{Z}$.
This analysis can also be extended for the synchrotron oscillation mode
and the first-order spin resonance $\nu_0\pm\nu_z=k, k \in \mathbb{Z}$,
obtaining a result similar to Eq.~(\ref{eq:first-order_resonance_strength})
and Eq.~(\ref{eq:first-order_resonance_zeta}) with $r$ replaced by $z$.

In ideal planar rings, 
$\vec{k}_0$ is in the horizontal plane, so that $\vec{\omega}_x\cdot\vec{k}_0$ and $\vec{\omega}_z\cdot\vec{k}_0$ vanish. Then the strengths
of first-order ``parent'' spin resonances $\nu_0\pm \nu_x=k, k \in \mathbb{Z}$
and $\nu_0\pm \nu_z=k, k \in \mathbb{Z}$
vanish too.
In contrast,
$\vec{\omega}_y\cdot\vec{k}_0 \approx
R(1+a\gamma)\frac{\frac{\partial B_x}{\partial y}}{B\rho}\vec{e}_x\cdot\vec{k}_0$ makes a nonzero
contribution to the strength of
the first-order ``parent'' 
spin resonances $\nu_0\pm \nu_y=k, k \in \mathbb{Z}$, which are
also called the ``intrinsic resonances''. 
 
Let's denote $K=\nu_0=k\pm\nu_y$ as the location of the intrinsic resonances, whose strength
$\tilde{\epsilon}_{K}^{\mathrm{intr}}$ can
be expressed as a function of the vertical betatron action $I_y$
\begin{eqnarray}    \tilde{\epsilon}_{K}^{\mathrm{intr},\pm}(I_y)&\approx&\frac{R(1+K)}{4\pi}\int_0^{2\pi} \frac{\frac{\partial B_x}{\partial y}}{B\rho} \sqrt{2I_y\beta_y} \nonumber \\
    &\times& e^{ i[K \Phi(\theta') \mp\nu_y\phi_y(\theta')\mp\psi_{y0}]}d\theta'
    \label{eq:intrinsic_resonance_strength}
\end{eqnarray}
where we use the relations $\vec{e}_x(\theta') \cdot \hat{k}_{0}(\theta')
\approx e^{i \nu_0 \Phi(\theta') }$ and $K=\nu_0=k\pm\nu_y$.
This also agrees with the result in Ref.~\cite{courant_acceleration_1980, leeSpinDynamicsSnakes1997}.

In realistic planar rings, 
the complex strength of these first-order spin resonances
are affected by the machine imperfections. In the context of
polarized beam acceleration, the intrinsic resonances
are generally more detrimental in comparison to other first-order spin resonances.

\subsection{The spin resonance structure of a model ring
\label{sec:model_ring}}

S. Y. Lee derived simple analytical formulas to explain the $K$-dependence of 
the strengths of the intrinsic resonances
and the imperfection resonances~\cite{lee_simple_1986,leeSpinDynamicsSnakes1997}, 
and thereby help to reveal interesting features of the spin-resonance spectrum. Most notably, 
it was shown that
the maximum resonance strength is determined by
two enhancement factors due to the regular arc FODO cells and the periodicity
of the machine, respectively. 

Here, we follow these analyses~\cite{lee_simple_1986,leeSpinDynamicsSnakes1997}
to study the structure of spin resonances for a model ring lattice with features shared with
very large electron rings, and discuss the general trend of evolution of
resonance strength with energy ($G\gamma$), to pave the way for the analysis
of the CEPC booster lattice.

\begin{figure}[!htb]
   \centering
   \includegraphics*[width=0.7\columnwidth]{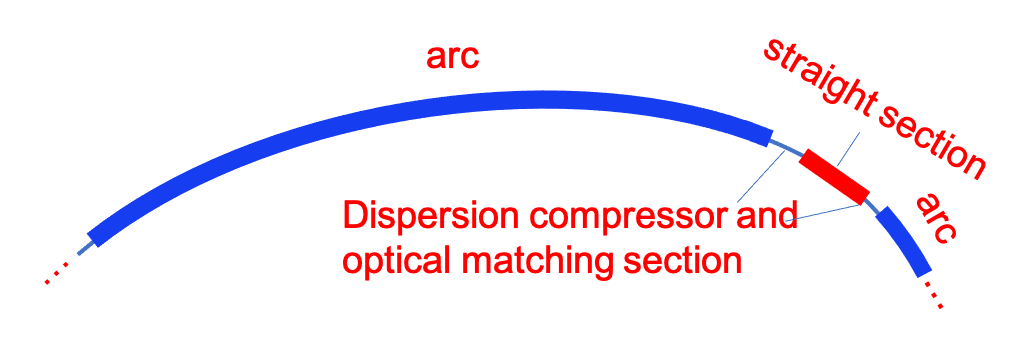}
   \caption{Diagram of the structure of the model ring, one of $P$ superperiods of the ring is showed.}
   \label{fig:detailed_structure}
\end{figure}
 
Let's consider a model ring lattice
composed of P 
superperiods,
as depicted in Fig.~\ref{fig:detailed_structure}.
Each superperiod includes an arc section containing $M$ identical FODO cells with dipoles, and 
a dispersion-free straight section containing $M'$ identical
FODO cells without dipoles. 
Each straight section and an adjacent arc section are 
connected by a dispersion suppressor and optical matching~(DOM) section.
Each superperiod contains two such 
sections in a mirror-symmetric layout. 
The vertical betatron phase advance of each arc FODO cell is denoted by $\phi_y^{\mathrm{cell}}=2\pi\mu$.
Then the total vertical betatron phase advance in all FODO cells in the arc regions amounts to
$2\pi\nu_B=2\pi MP\mu$.
The bending angles $\tilde \phi$ of dipoles in each arc FODO cell and each DOM section are $\tilde \phi^{\mathrm{cell}}$
and $\tilde \phi^{\mathrm{dis}}$, respectively. 
Let's denote the total 
bending angle in all arc sections as  $2\pi\eta_{\mathrm{arc}}$, then $\eta_{\mathrm{arc}}=\frac{M\tilde \phi^{\mathrm{cell}}}{M\tilde \phi^{\mathrm{cell}}+2\tilde \phi^{\mathrm{dis}}}$.

In addition, we assume that
the vertical betatron phase advance in one straight section is $2\pi L, L \in \mathbb{Z}$.
We also assume $P\ll M$ and $M' \ll M$, 
as in ultra-high energy rings like the CEPC booster.
Thin lens approximation for quadrupole and dipole fields are adopted so that approximations to 
the strength of the imperfection and the intrinsic spin resonances can be easily obtained for this model lattice
layout.

\subsubsection{The structure of intrinsic resonances}
\begin{figure}[hbt!]
\centering
\subfigure{
\begin{minipage}{\columnwidth}
\centering
\includegraphics[width=0.8\columnwidth]{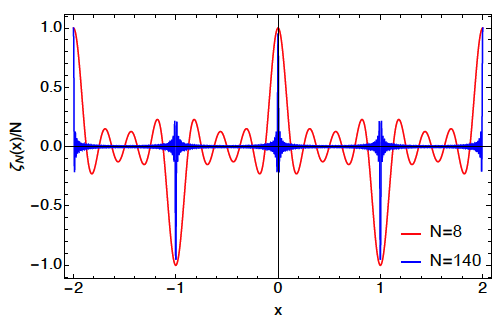} \\
\end{minipage}
}
\subfigure{
\begin{minipage}{\columnwidth}
\centering
\includegraphics[width=0.8\columnwidth]{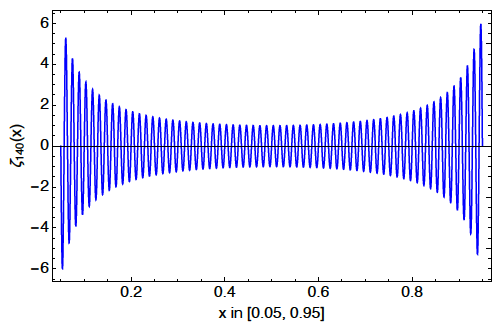} \\
\end{minipage}
}
\caption{The shape of the enhancement function $\zeta_N(x)$. The top plot shows 
$\zeta_8(x)/8$ and $\zeta_{140}(x)/140$ in the range $x\in [-2,2]$, the bottom plot shows 
$\zeta_{140}(x)$ in the range of $x \in [0.05,0.95]$.}
\label{fig:zeta_function}
\end{figure}

The strength of intrinsic resonances for this model ring can be regarded
as the contribution from a superperiod $\tilde\epsilon_{K,0}^{ \mathrm{intr}, \pm}$
multiplied by an enhancement factor $E_P^\mp$,
\begin{eqnarray}
    \tilde\epsilon_K^{ \mathrm{intr}, \pm} &=& \tilde\epsilon_{K,0}^{ \mathrm{intr}, \pm} \times E_P^\mp \nonumber \\
    E_P^\mp &=& e^{i\pi(P-1)\frac{K\mp\nu_y}{P}}\zeta_P(\frac{K\mp\nu_y}{P})
\end{eqnarray}
which contains an enhancement function $\zeta_N(x)$
depending on an integer $N$ and a real number $x$,
\begin{equation}
    \zeta_N{(x)}=\frac{\sin N\pi x}{\sin \pi x}
\end{equation}
As shown in Fig.~\ref{fig:zeta_function},
$\zeta_N(x)=0$ when $N*x\in\mathbb{Z}$ while $x \notin \mathbb{Z}$, and
$|\zeta_N(x)|\rightarrow {N}$, as $x$ approaches an integer $k$. In addition,
for a much larger $N=140$ relative to $N=8$ in Fig.~(\ref{fig:zeta_function}), the peaks are much narrower, and
the amplitude of $\zeta_N(x)$ decays more substantially for $x$ distant from integers. Additionally, as
illustrated in the bottom plot of 
Fig.~(\ref{fig:zeta_function}), 
for a large $N=140$,
when $x$ deviates 
slightly from an integer
by 0.05, the amplitude of $\zeta_N(x)$
is bounded within 6, less than $1/20$ of $N$,
indicating cancellation rather than enhancement among
the contributions of $N$ identical units.

According to the form of $\zeta_N(x)$, 
the enhancement factor $E_P^\mp$ is nonzero 
only for the intrinsic resonances $K=nP\pm \nu_y, n\in \mathbb{Z}$,
when the contributions from all the superperiods add up coherently, and $|E_P^\mp|=P$.

The contribution from one superperiod can be written as a sum of
three parts,
\begin{equation}
    \tilde\epsilon_{K,0}^{ \mathrm{intr}, \pm} = \tilde\epsilon^{\mathrm{intr},\pm}_{K,0,\mathrm{arc}}+\tilde\epsilon^{\mathrm{intr},\pm}_{K,0,\mathrm{str}}+\tilde\epsilon^{\mathrm{intr},\pm}_{K,0,\mathrm{DOM}}
\end{equation}
where $\tilde\epsilon^{\mathrm{intr},\pm}_{K,0,\mathrm{arc}}$,
$\tilde\epsilon^{\mathrm{intr},\pm}_{K,0,\mathrm{str}}$ and
$\tilde\epsilon^{\mathrm{intr},\pm}_{K,0,\mathrm{DOM}}$
represent the contributions from one arc section,
one straight section, as well as two DOM sections, respectively. 
The first two terms involve
summations of contributions from many identical FODO cells
and thus can be greatly simplified analytically, featuring
special patterns of enhancement or cancellation under specific
conditions. On the other hand, the third term depends on the specific
optics design and we are not aware of any 
general method for simplifying it.
However, the DOM sections typically cover
only a small fraction of the entire lattice, 
in particular in larger rings. In addition, we also find
a pattern whereby these contributions tend to cancel out
for small $K$.

Firstly, the contribution from one arc section can be regarded
as that from a standard arc FODO cell multiplied by an
enhancement factor $E_M^\mp$ due to the M FODO cells.
The simplified expression for
$\tilde\epsilon^{\mathrm{intr},\pm}_{K,0,\mathrm{arc}}$ is
\begin{eqnarray}
\label{eq:intrinsic_arc} 
\tilde\epsilon^{\mathrm{intr},\pm}_{K,0,\mathrm{arc}} &\approx& \frac{1+K}{4 \pi}\sqrt{2I_y} (g_f +  g_de^{i\frac{K \eta_{\mathrm{arc}}\mp\nu_B}{MP}\pi}) E_M^\mp  \nonumber \\ 
E_M^\mp &=& e^{i\pi(M-1)\frac{K\eta_{\mathrm{arc}}\mp\nu_B}{PM}}\zeta_M(\frac{K\eta_{\mathrm{arc}}\mp\nu_B}{PM})
\end{eqnarray}
where 
\begin{equation}
g=\frac{R}{B\rho}\int_{\mathrm{quad}} \frac{\partial B_x}{\partial y}\sqrt{\beta_y}d\theta
\label{eq:g_factor_1}
\end{equation}
is an integral over a quadrupole and $g_f$ and $q_d$ denote
the corresponding values for the focusing and defocusing quadrupoles in an arc FODO cell, respectively. 
According to the properties of $\zeta_N(x)$,
the amplitude of the enhancement factor 
$|E_M^\mp|$ approaches $M$ near $K=(mPM \pm \nu_B)/\eta_{\mathrm{arc}}, m\in \mathbb{Z}$,
when the contribution from all the FODO cells in an arc add up coherently. However, 
$|E_M^\mp|$ reduces to
a small fraction of $M$ when $K$ deviates from these
conditions by a corresponding $\Delta K$.
For example, if we take $P=8$, $M=140$,
and $\eta_{\mathrm{arc}}=140/142$,
then $|E_M^\mp|<6$
for $\Delta K \ge 56.8$.

Next, the contribution from one straight section
can also be regarded as that of a standard FODO cell 
without dipoles multiplied
by an enhancement factor $E_{M'}^{'\mp}$ due to
the $M'$ FODO cells to give
\begin{eqnarray}  
\tilde\epsilon^{\mathrm{intr},\pm}_\mathrm{K,0,str}&\approx&\frac{1+K}{4 \pi}\sqrt{2I_y} (g_f'+g_d'e^{i\frac{\mp\nu_{\mathrm{str}}}{PM'}\pi})   E_{M'}^{'\mp}  \nonumber \\
E_{M'}^{'\mp}&=&e^{i\pi(M'-1)\frac{\mp\nu_{y,\mathrm{str}}}{PM'}}\zeta_{M'}(\frac{\mp\nu_{y,\mathrm{str}}}{PM'})
    \label{eq:intrinsic_straight_1}
\end{eqnarray}
where $g_f'$ and $g_d'$ denote the integral defined in Eq.~(\ref{eq:g_factor_1}) for the focusing and defocusing
quadrupoles in a standard FODO cell of the straight sections, respectively. 
$2\pi\nu_{y,\mathrm{str}}$ is the total
vertical betatron phase advance in all straight sections.
Since we assume the vertical betatron phase advance in one straight section is $2\pi L, L \in \mathbb{Z}$,
$\nu_{y,\mathrm{str}}=PL$ for the model ring.
According to the properties of $\zeta_N(x)$, when $M' \neq L$,
$E_{M'}^{'\mp}$ is zero and the 
contribution from straight sections to the strength of intrinsic resonances vanishes.

Now, we are ready to summarize the features of the strengths of intrinsic resonances
for this model ring.
The straight sections have zero
contribution to the resonance strength, while the DOM sections cover
only a small fraction of the whole lattice. Therefore,
when $|E_{P}^{\mp}|$ equals to $P$ and $|E_{M}^{\mp}|$ is close to the maximum values, 
so that the contributions from all $PM$ arc FODO cells add up,
the resonance strengths are greatly enhanced. 
We call these resonances the super-strong intrinsic resonances. As $P \ll M$,
this occurs at those $K=nP \pm \nu_y$ which are closest to $(mPM \pm \nu_B)/\eta_{\mathrm{arc}}, m\in \mathbb{Z}$.

For a large $M$,
away from these super-strong intrinsic resonances, the amplitude
of $E_{M}^{\mp}$ decays rapidly, making the contributions from
arc sections much weaker
and those from DOM sections more significant.
Additionally,
there can be cancellations between the contributions 
from the arc sections and the DOM sections, 
depending on the lattice parameters. In particular,
when $\nu_y$ is large and $K\ll \nu_y$, 
the exponential factor
$e^{ i[K \Phi(\theta') \mp\nu_y\phi_y(\theta')\mp\psi_{y0}]}$
in Eq.~(\ref{eq:intrinsic_resonance_strength}) includes a fast
wave with a phase $\nu_y\phi_y(\theta')$ modulated by the slow wave with a phase $K\Phi(\theta')$, so that the contributions from all
FODO cells in each superperiod tend to cancel out. 
Such a cancellation generally becomes more incomplete as
$K$ increases.

Our analysis indicates a general trend for the strengths of intrinsic resonances.
For a fixed vertical betatron amplitude $I_y$,
as $K$ increases, we generally
expect that the
resonance strength increases until the first super-strong resonance is reached,
after which it oscillates as adjacent super-strong resonances
are approached and left behind.

In practical lattices, the straight sections
may not be composed of identical FODO cells and may not
have the assumed
betatron phase advance.
However, efficient
cancellation can still occur among a large number of quadrupoles as the
phase $K\Phi(\theta')$ remains almost constant in straight sections. 
Additionally, the unique functions of each straight section
and machine-geometry considerations
such as bypasses near the interaction regions,
can break lattice periodicity,
resulting in some relatively 
weak spin resonances different from the
condition $K=nP \pm \nu_y$. Nevertheless,
the method of determining
locations of super-strong resonances and the general trend for
the strengths of intrinsic 
resonances are still valid.

\subsubsection{The structure of imperfection resonances}
The expression of the strength of imperfection resonances,
in Eq.~(\ref{eq:imperfection_resonance_strength}),
includes the vertical closed orbit $y_{\mathrm{co}}$,
which is related to the radial magnetic field errors $(\frac{\Delta B_x}{B\rho})_0$ via the Hill's
equation~\cite{leeSpinDynamicsSnakes1997}
\begin{equation}
\frac{d^2(y_{\mathrm{co}}/\sqrt{\beta_y})}{d\phi_y^2}+\nu_y^2 (y_{\mathrm{co}}/\sqrt{\beta_y})=\nu_y^2\beta_y^{3/2}(\frac{\Delta B_x}{B\rho})_0
\end{equation}
the solution for $y_{\mathrm{co}}$ is
given by
\begin{equation}
    y_{\mathrm{co}}(\theta)=\beta_y(\theta)^{1/2}\sum_{k=-\infty}^\infty \frac{\nu_y^2 f_k e^{ik\phi_y(\theta)}}{\nu_y^2-k^2}
    \label{eq:closed_orbit_harmonic}
\end{equation}
where $f_k$ is the Fourier amplitude of the error harmonic $k$ given by
\begin{equation}
    f_k=\frac{R}{2\pi\nu_y}\oint \sqrt{\beta_y  }(\frac{\Delta B_x}{B\rho})_0 e^{-i k \phi_y(\theta)}d\theta
\end{equation}
The strengths of imperfection resonances can then be evaluated as~\cite{leeSpinDynamicsSnakes1997}
\begin{eqnarray}
    {\tilde \epsilon}_K^{\textrm{imp}}&\approx&
    -\frac{R(1+K)}{2\pi }\sum_{k=-\infty}^\infty\frac{\nu_y^2 f_k }{\nu_y^2-k^2} \nonumber \\
    &\times& 
    \oint \frac{\frac{\partial B_x}{\partial y}}{B\rho} \beta_y^{1/2}  e^{i(k\phi_y+ K \Phi)}d \theta
    \label{eq:imperfection_resonance_strength_1}
\end{eqnarray}


The strengths of imperfection resonances
of this model ring can also be written
as sums of three contributions from the arc sections, the straight sections
and the DOM sections, respectively,
\begin{equation}
    \tilde\epsilon_K^{ \mathrm{imp}} = \tilde\epsilon^{\mathrm{imp}}_{K,\mathrm{arc}}+\tilde\epsilon^{\mathrm{imp}}_{K,\mathrm{str}}+\tilde\epsilon^{\mathrm{imp}}_{K,\mathrm{DOM}}
\end{equation}
The first two terms involve
summations of the contributions from many identical FODO cells
of different superperiods
and thus can be greatly simplified analytically, and feature
special patterns of enhancement or cancellation under certain conditions.

Firstly, the contribution from all arc sections can be expressed as~\cite{leeSpinDynamicsSnakes1997}
\begin{eqnarray}
\label{eq:imperfection_resonance}
    \tilde\epsilon^{\mathrm{imp}}_{K,\mathrm{arc}} &\approx& \frac{1+K}{2 \pi}\sum_{k=-\infty}^{\infty}\frac{\nu_y^2 f_k}{\nu_y^2-k^2}e^{i\frac{P-1}{P}(k+K)\pi} \nonumber \\
    &\times&e^{i\frac{M-1}{PM}(    K \eta_{\mathrm{arc}} +\frac{k\nu_B}{\nu_y}      )\pi} \Bigl[g_d 
    -  g_fe^{-i\frac{K \eta_{\mathrm{arc}}+\frac{k\nu_B}{\nu_y}}{MP}\pi} \Bigr] \nonumber \\
        &\times&\zeta_P(\frac{k+K}{P}) \zeta_M(\frac{K \eta_{\mathrm{arc}}+k\frac{\nu_B}{\nu_y}}{MP})
\end{eqnarray}
For a specified $k$, $|\zeta_P|=P$ for
$K=nP-k, n\in \mathbb{Z}$, so that the contributions
from all superperiods add up coherently. If this condition is not met,
$\zeta_P=0$.
$|\zeta_M|$ approaches
$M$ as $K$ approaches $(mPM-k\frac{\nu_B}{\nu_y})/\eta_{\mathrm{arc}}, m\in \mathbb{Z}$, so that the contributions from all FODO cells
in an arc section add up coherently. 
Note that for each integer value of $K$,
a family of $k$ always exists that satisfies
$K=nP-k, n\in \mathbb{Z}$, so that 
there is no complete cancellation in the strength of
imperfection resonances due
to the lattice periodicity, unlike the case for intrinsic resonances.
This can be viewed as the result of symmetry breaking due to
machine imperfections.

In addition, the strengths of
imperfection resonances depend on the spectrum of $f_k$.
For an imperfect lattice before dedicated closed orbit
correction, the most important terms are 
those with $|k|$ near $\nu_y$,
as the factor $|\frac{\nu_y^2 }{\nu_y^2-k^2}|$
can be very large. In particular, let's consider the terms
$k=\pm [\nu_y]$ where $[x]$ denotes the integer nearest to
a real number $x$, the
corresponding super-strong imperfection resonances
are those when both conditions
$K=mP\pm[\nu_y]$ and $K=[(nPM\pm[\nu_y]\frac{\nu_B}{\nu_y})/\eta_{\mathrm{arc}}]$ are nearly satisfied. 

Note that the closed-orbit correction can significantly 
change the spectrum of $f_k$,
and tends to reduce the amplitudes of $f_k$ with
$|k|$ near $\nu_y$ as well as other harmonics,
in a way depending on the detailed correction
algorithms and settings.
Even if the super-strong imperfection resonances as
described above can
become less prominent after the closed
orbit correction, they are likely to be surrounded by 
plateaus of strong imperfection resonances.

Second, the contribution from all straight sections can
also be regarded as a sum over harmonics $k$: 
\begin{eqnarray}
    \tilde\epsilon^{\mathrm{imp}}_{K,\mathrm{str}} &\approx& \frac{1+K}{2 \pi}\sum_{k=-\infty}^{\infty}\frac{\nu_y^2 f_k}{\nu_y^2-k^2} 
    e^{i\frac{P-1}{P}(k+K)\pi}\nonumber \\
    &\times&
    e^{i(M'-1)  \frac{\mp\nu_{y,\mathrm{str}}\frac{k}{\nu_y}}{PM'}      \pi}\Bigl[g_d 
    -  g_fe^{   -i   \frac{\mp\nu_{y,\mathrm{str}}\frac{k}{\nu_y}}{PM'}   \pi     } \Bigr] \nonumber \\
    &\times&\zeta_P(\frac{k+K}{P})\zeta_{M'}(\frac{\mp\nu_{y,\mathrm{str}}\frac{k}{\nu_y}}{PM'} )     
\end{eqnarray}
In this case, no matter what $\nu_{y,\mathrm{str}}$ is, there will always be many $k$ that make the enhancement factor $\zeta_{M'}$ nonzero, so that the total sum will not be zero. If $M'\neq L$ and $\nu_y$ is large, since $k=\nu_y$ is a zero point of $\zeta_{M'}$, for those most important items ($|k|$ near $\nu_y$), their corresponding $\zeta_{M'}$ will be close to zero so that the contribution of the straight section will 
be relatively small.

To summarize, the structure of imperfection resonances
with the contribution from only one harmonic $k$, is quite similar to the
structure of the intrinsic resonances.
Besides the peaks when the contributions from all arc FODO cells add up coherently,
there is also cancellation among all FODO cells if $k\gg 1$ and $K \ll k$.
Nevertheless, the strength of an imperfection resonance is the sum over various harmonics $k$ modulated by $\frac{\nu_y^2 f_k }{\nu_y^2-k^2}$,
with varying locations of enhancement, and thus
strongly depends on the spectrum of $f_k$. In general, after the closed-orbit correction, the $f_k$ terms with $|k|$ near $\nu_y$ become weaker, the terms with $|k|$ further away from $\nu_y$ are less reduced, forming a plateau around the original peak.
Generally, we expect that the strength of imperfection resonances increases with $K$ until
reaching the plateau near the first super-strong imperfection resonance, after which
it oscillates as adjacent super-strong imperfection resonances are approached and left behind.

\section{CEPC BOOSTER LATTICE SETUP\label{sec:booster_lattice}}

\begin{figure}[!htb]
   \centering
   \includegraphics*[width=0.7\columnwidth]{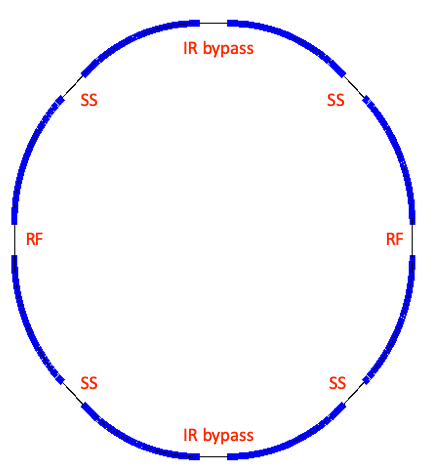}
   \caption{Layout of the candidate lattice for the
CEPC booster.}
   \label{fig:booster_layout}
\end{figure}

In this paper we study a simplified lattice for the CEPC
booster.
As shown in Fig.~\ref{fig:booster_layout}, the lattice is composed of 8 
identical arc
sections, 8 straight sections
and 16 DOM sections connecting them.
Each arc section 
consists of 140 standard FODO cells with
a length of 71~meters and 90 degree phase 
advances in both horizontal and vertical planes.
The straight sections include two bypass regions 
near the interaction regions of the collider rings
~(``IR bypass'' in Fig.~\ref{fig:booster_layout}), 
two RF regions that accommodate the RF cavities
~(``RF'' in Fig.~\ref{fig:booster_layout}), as well 
as four other regions for injection and extraction and 
other functionalities. However, the
optics design for these special functions is not
implemented in this lattice. Instead,
each of these straight sections 
contains 24 identical FODO cells with
the total horizontal and vertical betatron
phase advances both of $6*2\pi$.
Each DOM section includes a dispersion suppressor
and an optics-matching region,
consisting of 6 irregular FODO cells.
The dispersion suppressor
consists of two FODO cells with dipoles
that have half the
bending angle of normal arc dipoles.
The optics-matching region consists of
four FODO cells
that match the optics between the dispersion suppressor
and the straight section. 
The strengths of quadrupoles are the same in the optics-matching regions
next to the bypass regions and the RF regions, while quadrupoles in the optics-matching
regions next to the four other straight sections have been
 slightly adjusted simultaneously for fine tuning of the betatron tunes, effectively
 reducing the lattice periodicity to 4.
The circumference of the 
booster is 100016.4~m, the same as the collider rings,
of which 80\% is the arc region. 
In this paper, we discuss three operation modes of the booster
with the extraction energy at 45.6~GeV in the Z-mode, 80~GeV in
the W-mode and 120~GeV in the H-mode.
The main parameters of
the CEPC booster lattice at these three beam energies are listed in Table~\ref{tab:lattice_parameter}.

\begin{table}[b]
\caption{Parameters of the CEPC booster lattice}
\begin{ruledtabular}
\begin{tabular}{lcdr}
\textrm{Operation mode}&
\textrm{Z}&
\multicolumn{1}{c}{W}&
\textrm{H}\\
\colrule
  Extraction energy~(GeV)       &45.6 &80.0    &120.0       \\ 
  $G\gamma$ at  extraction energy     &103.48 &181.55    &272.33      \\ 
           Betatron tune $\nu_x/\nu_y$ &   \multicolumn{3}{c}{353.18/353.28}   \\ 
           Synchrotron tune $\nu_z$ &0.1  &0.1&0.13 \\
           Natural emittance $\epsilon_{\textrm{natural}}$~(nm)       & 0.19     & 0.58  &1.29    \\
           Transverse damping time(s) & 0.878&0.162 &0.048 \\
           Vertical damping time(s) & 0.889&0.164 &0.048 \\
           Longitudinal damping time(s) & 0.448& 0.082 &0.024 \\
\end{tabular}
\end{ruledtabular}
\label{tab:lattice_parameter}
\end{table}

To reflect the influence of machine imperfections on
beam polarization, we introduced misalignment errors and relative field errors of magnets,
listed in Table~\ref{tab:error_setup}, into the lattice. 
Using the Accelerator Toolbox~(AT) code~\cite{at},
these errors
were generated according to a Gaussian distribution truncated at $\pm 3\sigma$. We then performed closed-orbit correction and betatron-tune correction for these
error seeds to restore the lattice performance. 
These error seeds were then converted to the lattice format of Bmad~\cite{Sagan:Bmad2006,Bmad}, for the calculations of closed
orbit and optics parameters, which were then used to calculate
the strength of imperfection resonances.
Finally, we also used Bmad for detailed multi-particle simulations of orbital and spin motion.

\begin{table}[!htb]
\caption{Magnet error settings }
{\begin{tabular}{@{}l|cccc|c@{}} \hline \hline
\multirow{2}*{Component}&\multicolumn{4}{c|}{Misalignment error}&\multirow{2}*{Field error} \\ \cline{2-5}
~&$\Delta x (\si{\um})$&$\Delta y (\si{\um})$&$\Delta z (\si{\um})$&$\Delta \theta_z (\si{\micro\radian})$&~\\ \hline
Dipole&100&100&100&100&0.05\% \\
Quadrupole&100&100&100&100&0.02\% \\
Sextupole&100&100&100&100&0.03\% \\
\hline \hline
\end{tabular} \label{tab:error_setup}}
\end{table}

For the closed-orbit correction, we imagined that there would be extra windings on the focusing and defocusing quadrupoles and used them as horizontal and vertical orbital correctors, respectively.  Moreover, we assumed that BPMs are attached to all the quadrupoles, four per betatron wave period. Note that these are reasonable approximations in this preliminary analysis before more realistic modeling of the correction procedure with stand-alone correctors and BPMs
is carried out.
There are inevitably some differences
between the magnetic centers of quadrupoles and the
electric centers of adjacent BPMs, namely the BPM offsets,
and this has substantial influence on the residual closed
orbit distortion~(COD) after correction.
As will be shown later, and as expected,
the level of COD affects
the polarization loss during the acceleration.
At this stage, we are not in the position to predict
what level of BPM offset or COD can be achieved
for the CEPC booster. Instead, we try to explore
the dependence of polarization loss on the
level of COD.
To this end, we introduced BPM offsets that also follow
a Gaussian distribution truncated at $\pm 3\sigma$.
We scanned a range of the rms BPM offset from 30~\si{\um} to 180~\si{\um},
with an interval of 30~\si{\um}.
For each setting of rms BPM offset,
we generated error seeds and carried out
closed-orbit correction using the singular value
decomposition~(SVD) algorithm.
Note that the closed-orbit correction to 
some error seeds failed to converge so that no closed
orbit was found.
These failed
error seeds were simply removed.
We are aware that
the algorithm of closed-orbit correction can be improved by
adjusting the truncation of the singular values and
employing multiple iterations, but this is beyond
the scope of this paper.
For each setting of rms BPM offset, we accumulated 10 error seeds with
converged closed-orbit correction, and then corrected the betatron tunes
using the quadrupole knobs in the DOM sections without affecting the arc and straight sections. 
Then we obtained a collection of 60 error seeds
for evaluation of the attainable polarization transmission.

\begin{figure}[!htb]
   \subfigure[Horizontal rms CODs]{
    \begin{minipage}{\columnwidth}
    \centering
    \includegraphics*[width=1\columnwidth]{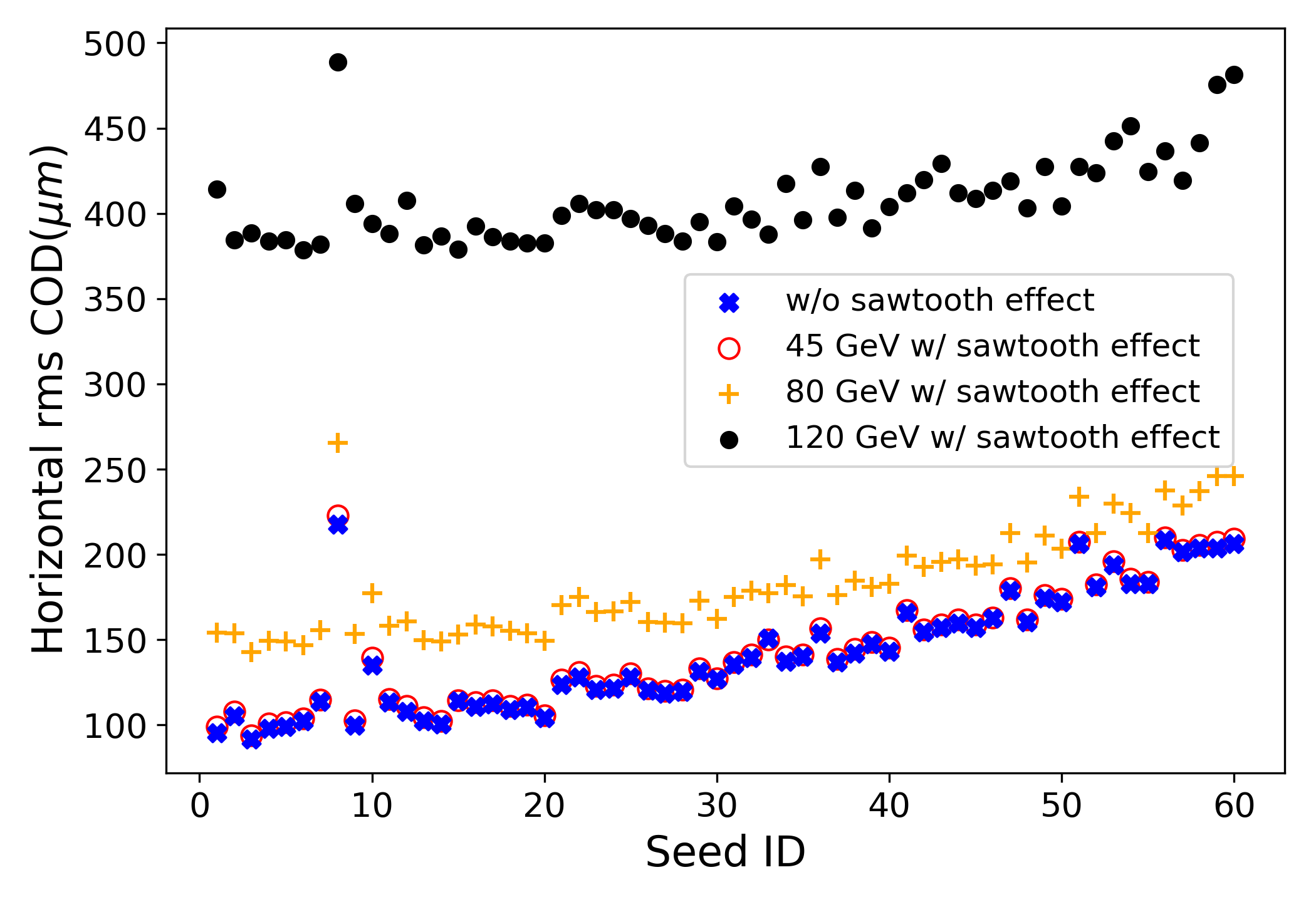} \\
\end{minipage}
}
   
   \subfigure[Vertical rms CODs]{
    \begin{minipage}{\columnwidth}
    \centering
   \includegraphics*[width=1\columnwidth]{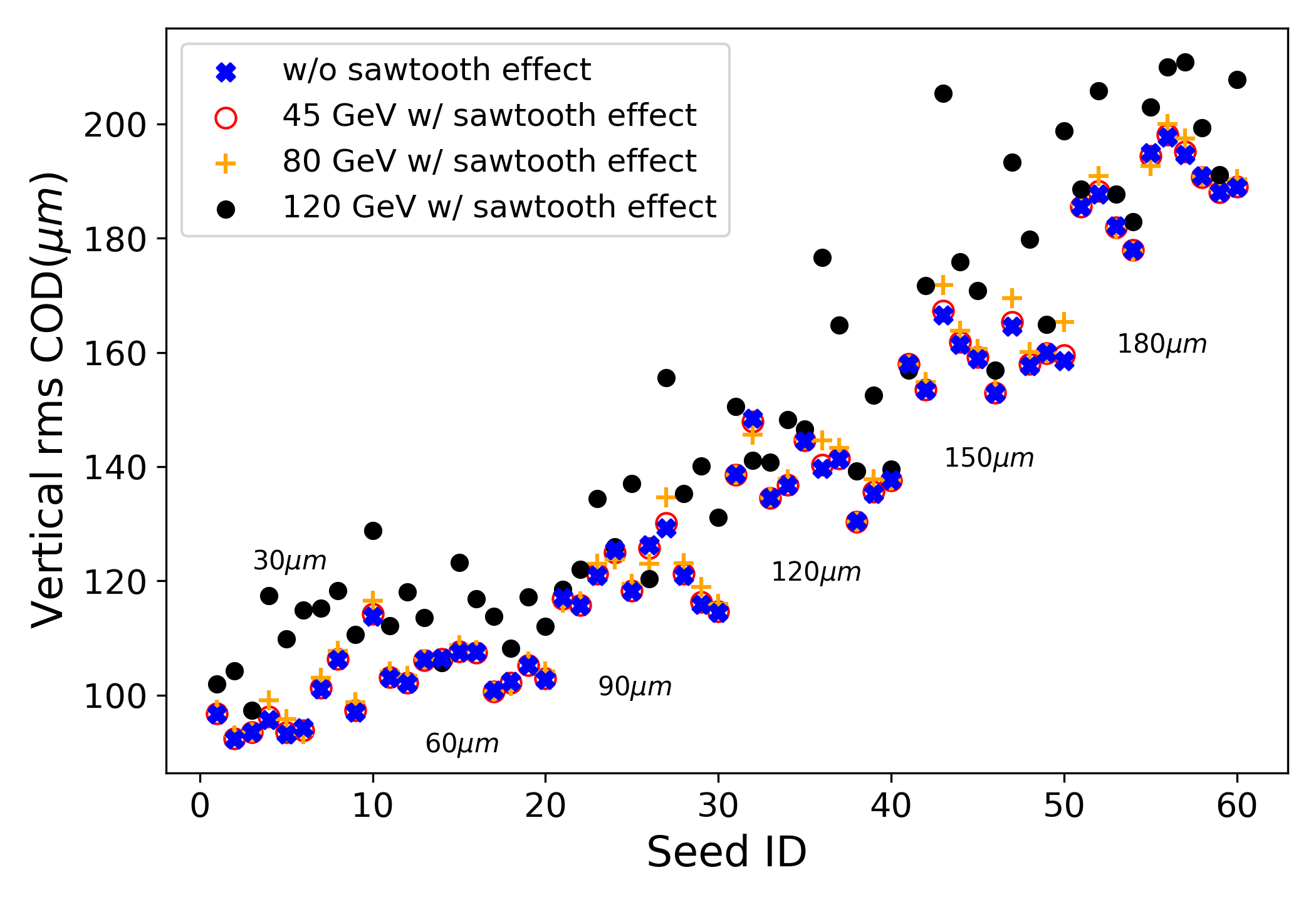} \\
\end{minipage}
}

   \caption{Rms CODs of the 60 error seeds after correction
   for the case without the sawtooth effect as well as the cases with the sawtooth effect
   at three different beam energies, the top and bottom
   plots show the rms CODs in the horizontal and vertical planes, respectively. The rms BPM offsets for each subset of error seeds are marked in the bottom plot.
 }
   \label{fig:seed_yco}
\end{figure}

In electron circular accelerators,
the element-by-element synchrotron-radiation energy loss accumulates in the 
arc sections until it is compensated in the RF cavities localized in
straight sections. This leads to an ``energy sawtooth'' for the reference particle,
and a sawtooth-shape contribution to the CODs in dispersive regions, 
leading to optics perturbations.
Hereafter we refer this effect as the sawtooth effect.
In the above correction procedures,
the element-by-element synchrotron-radiation energy loss was turned
off, so that the sawtooth effect was not taken into account. 
Fig.~\ref{fig:seed_yco} shows 
the horizontal and
vertical rms CODs of the 60 error seeds for the case without 
the sawtooth effect, as well as the cases at a beam energy of 45.6~GeV, 80~GeV and 120~GeV 
with the sawtooth effect.
The rms vertical CODs in the
case without the sawtooth effect are in the range of 80~\si{\um} to 220~\si{\um}.
There is a general trend that the rms
vertical COD correlates with the rms BPM offset, apart from 
a few outliers.
Since the amplitude of the energy sawtooth scales with $E^3$, the sawtooth
effect is much more pronounced at higher beam energies.
Fig.~\ref{fig:seed_yco} shows that
the horizontal and vertical rms CODs of the case at 45.6~GeV with
the sawtooth effect
are very close to the results of the case without the sawtooth effect, while the difference between
the cases with and without the sawtooth effect becomes more substantial at 80~GeV and 120~GeV.
Since the amplitude of the vertical dispersion 
is generally much smaller than the horizontal dispersion, the influence of the sawtooth effect on the vertical
rms COD is less severe relative to that on the horizontal rms COD. 
As we will elaborate
later in this paper, this difference in the vertical rms COD also affects
the calculation of the strength of imperfection resonances
and thus the estimation of the
depolarization during the acceleration.

\begin{figure}[!htb]

\subfigure[Vertical equilibrum emittance and the ratio between transverse equilibrium emittances at 120~GeV]{
\begin{minipage}{\columnwidth}
\centering
\includegraphics[width=1\columnwidth]{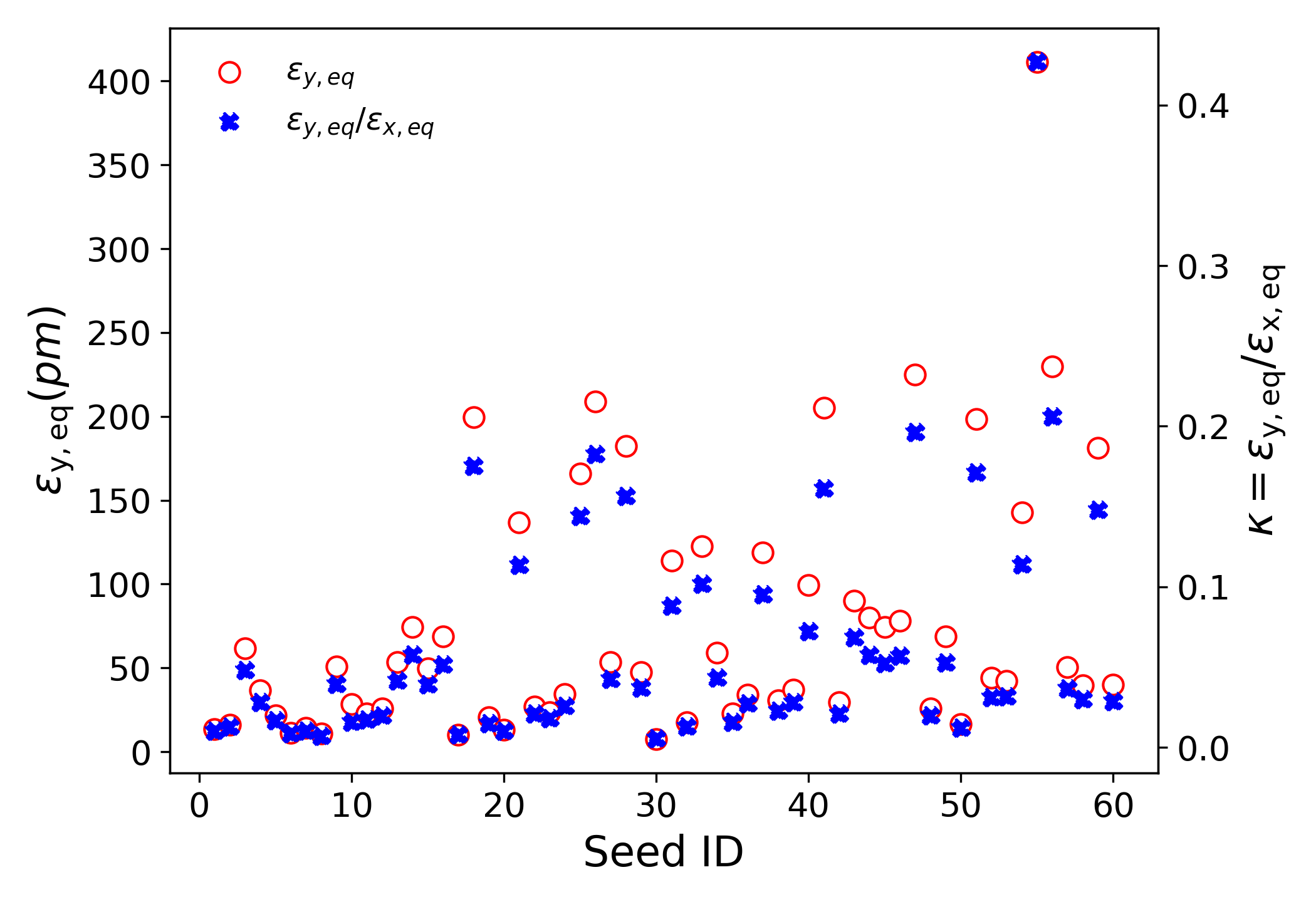} \\
\end{minipage}
}

\subfigure[Ratio of the vertical equilibrum emittance at different energies to that at 10~GeV]{
\begin{minipage}{\columnwidth}
\centering
\includegraphics[width=1\columnwidth]{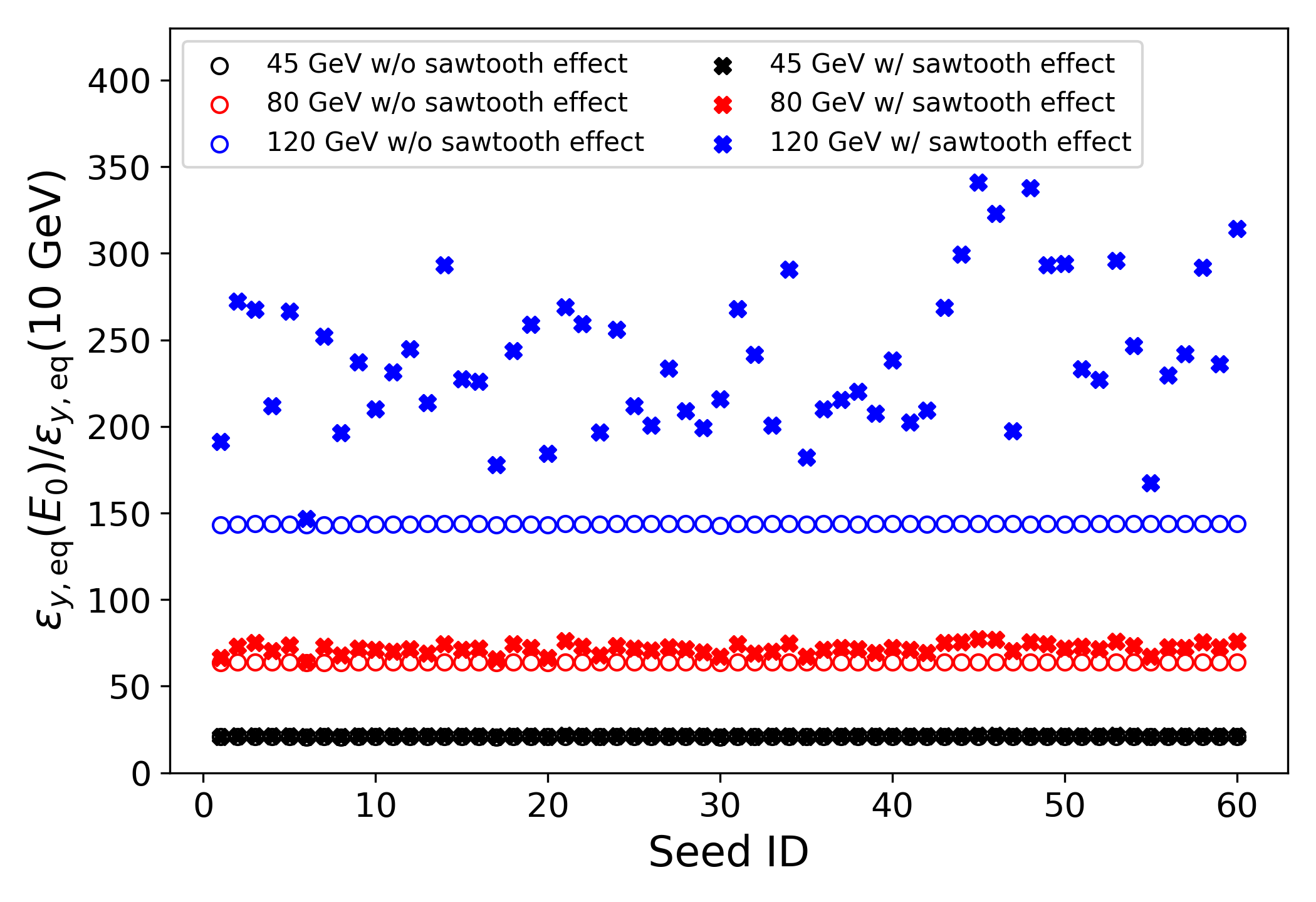} \\
\end{minipage}
 }   
   
   \caption{The equilibrium emittances of the 60 error seeds.
   (a) shows the vertical equilibrum emittance and the ratio between transverse equilibrium emittances at 120~GeV. (b) shows the ratio of the vertical equilibrium emittance
   at different energies relative to that at 10 GeV, for the cases
   without and with sawtooth effect, respectively.
   }
   \label{fig:seed_emity}
\end{figure}

Fig.~\ref{fig:seed_emity}(a) shows the vertical equilibrium
emittances $\epsilon_{y,\mathrm{eq}}$ and the ratios between the transverse equilibrium emittances
$\kappa=\epsilon_{y,\mathrm{eq}}/\epsilon_{x,\mathrm{eq}}$,
for the 60 error seeds at a beam energy of 120~GeV,
calculated using the beam envelope formalism~\cite{ohmi_beam-envelope_1994} implemented in Bmad,
which takes into account of the sawtooth effect. 
$\epsilon_{y,\mathrm{eq}}$ is in the range of 6~pm to about 500~pm, and
$\kappa$ lies between
0.49\% to 34.8\% among the error seeds. 
Dedicated optics corrections in particular the dispersion-free
steering~\cite{assmann_emittance_2000} and coupling correction~\cite{franchi_vertical_2011},
can in principle be
implemented to reduce $\kappa$ to a few percent or even
lower. Nevertheless, this study aims to show how the vertical
equilibrium emittance would affect the polarization transmission
in the booster. The necessity and the method of
implementing the low emittance tuning are beyond the scope of
this paper.
In addition, without the sawtooth effect,  
$\epsilon_{y,\mathrm{eq}}(E)$ is proportional to $E^2$,
while with the sawtooth effect, the change in the vertical
dispersion and transverse coupling would alter this
proportion.
Fig.~\ref{fig:seed_emity}(b) shows $\epsilon_{y,\mathrm{eq}}(E)/\epsilon_{y,\mathrm{eq}}(10~\mathrm{GeV})$
at the three extraction beam energies for the 60 error seeds, with and without the sawtooth effect,
respectively. The
 difference between the cases with and
without the sawtooth effect becomes clear
at higher beam energies above 80~GeV.

\begin{figure}[!htb]
   \centering
   \includegraphics*[width=1\columnwidth]{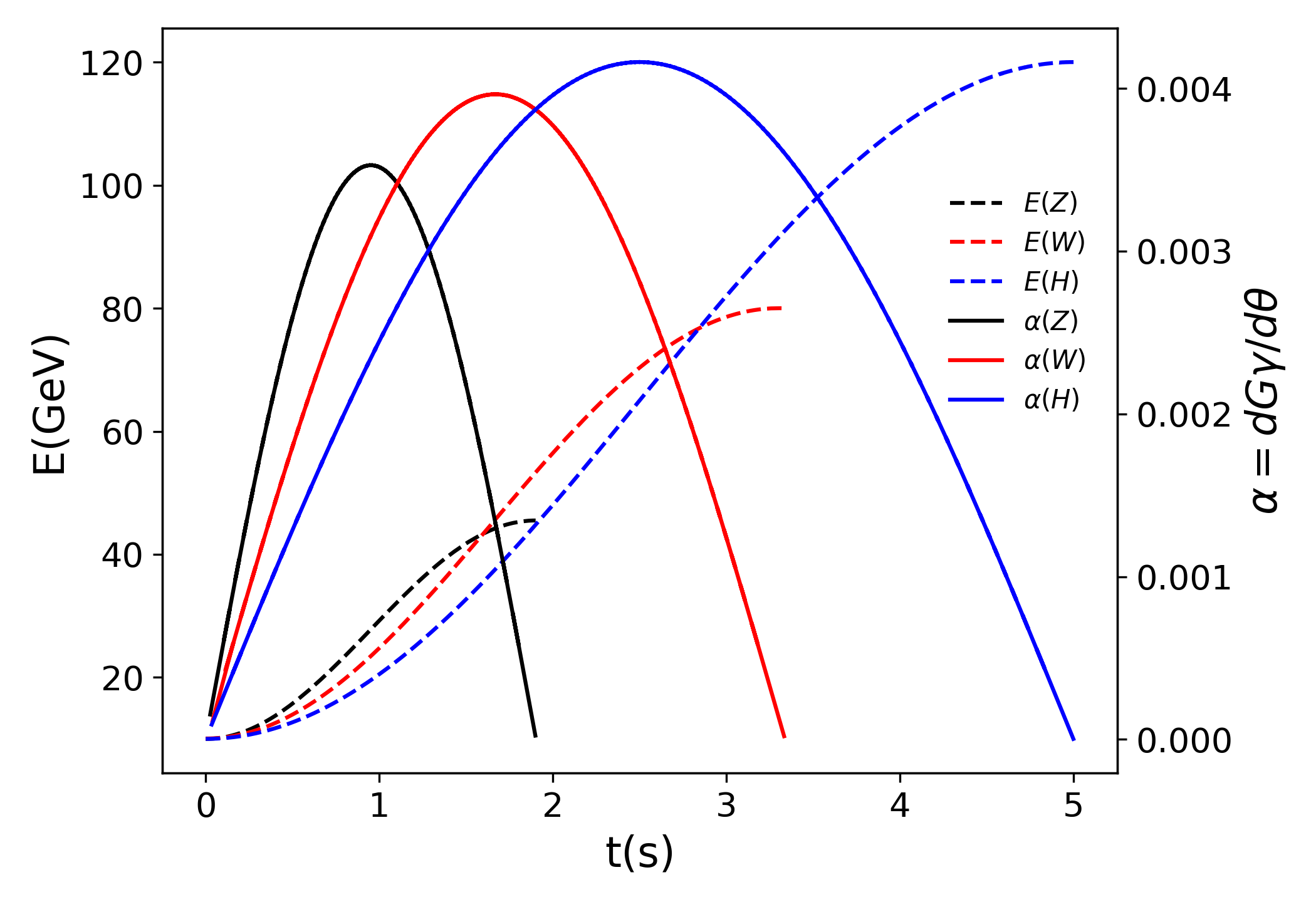}
   \caption{The evolution of the beam central energy (in dashed lines) and the spin-resonance crossing rate $\alpha$~(in solid lines) for the 3 different operation modes.}
   \label{fig:ramping_curve}
\end{figure}

The time dependence of the energy ramping is also very
relevant for the study of beam polarization evolution
in the booster. Following the CEPC CDR,
we adopted a cosine-shaped energy ramping curve
\begin{equation}
    E(t)=E_{\mathrm{inj}}+\frac{(E_{\mathrm{ext}}-E_{\mathrm{inj}})}{2}(1-\cos(\frac{\pi t}{t_{\textrm{ramp}}}))
    \label{eq:energy_ramp}
\end{equation}
With this, we fixed the booster injection energy to
10~GeV,
and set the ramping time $t_{\textrm{ramp}}$ to 1.9~s, 3.3~s and 5.0~s for the
acceleration to 45.6 GeV~(Z-mode), 80 GeV~(W-mode)
and 120 GeV~(H-mode), respectively.
The RF voltages and phases
were set to compensate for the synchrotron radiation 
energy loss during the ramp, as well as for maintaining a fixed
synchrotron tune, as listed for different extraction
energies in Table~\ref{tab:lattice_parameter}. 
Accordingly, 
for the bare lattice,
we can evaluate the spin tune on the design
orbit $G\gamma(t)$,
as well as the resonance crossing rate $\alpha(t)$,
\begin{eqnarray}
G\gamma(t)&=&\frac{E[\mathrm{GeV}](t)}{0.4406485} \nonumber \\
\alpha(t)&\approx&\frac{d(G\gamma(t))}{d\theta}
\end{eqnarray}
Note that the beam particles also
execute synchrotron oscillations in the acceleration process, which
would modify the resonance crossing rate,
in particular near the injection and extraction beam energies, when the energy ramping rate is lower.
Fig.~\ref{fig:ramping_curve} shows the evolution
of the beam energy and the resonance crossing rate 
during the acceleration process for the three different operation modes.
Due to the large circumference of the CEPC booster, the resonance crossing rate is above 0.001 most
of the time during the acceleration process, so that 
according to the Froissart-Stora formula, the crossings of spin resonances with $|\tilde \epsilon|<0.019$ 
are within the fast crossing regime.

Moreover, in the analysis and simulations of the evolution
of the beam polarization, we assume the injected beam 
has a six-dimensional Gaussian distribution,
with a transverse distribution matched to the
Courant-Snyder
parameters at the injection point of the booster
lattice. The injected beam parameters are listed in Table~\ref{tab:parameters_of_injection_beam}, where
the transverse rms emittances are more conservative
than the specifications outlined in the CEPC CDR.

\begin{table}[!htb]
\caption{Parameters of the booster injected beam}
\begin{ruledtabular}
\begin{tabular}{lcdr}

  Injection beam energy~(GeV)       &10.0       \\ 
  Rms energy spread~(\%)    &0.16     \\ 
  Rms bunch length~(mm) &1.0\\
Horizontal and vertical rms emittance~(nm) &80/40
          
\end{tabular}
\end{ruledtabular}
\label{tab:parameters_of_injection_beam}
\end{table}

After injection into the booster, the 
evolution of the vertical rms 
beam emittance $\epsilon_{y,\mathrm{rms}}(t)$
can be approximated following~\cite{lee_accelerator_2004}
as
\begin{eqnarray}
    \frac{d\epsilon_{y}^{N}(t)}{dt}&=&-\left[
    \frac{1}{E(t)}\frac{dE(t)}{dt}+\frac{2}{\tau_{y,0}}(\frac{E(t)}{E_0})^3\right]\epsilon_{y}^{N}(t) \nonumber \\
    &+&\frac{2}{\tau_{y,0}}(\frac{E(t)}{E_0})^5
    \label{eq:emittance_damp}
\end{eqnarray}
where $\epsilon_{y}^{N}=\epsilon_{y,\mathrm{rms}}(t)/\epsilon_{y,\mathrm{eq},0}$,
and $\epsilon_{y,\mathrm{eq},0}$ and $\tau_{y,0}$ are the vertical 
equilibrium emittance
and the radiation damping time at the
reference energy $E_0=120$~GeV, respectively.
This equation accounts for the 
combined effects of adiabatic damping, 
radiation 
damping and quantum excitation.
Including the sawtooth effect is not straightforward and is ignored here, though its influence is not necessarily
negligible, as illustrated by Fig.~\ref{fig:seed_emity}(b).
Given the initial vertical rms emittance,
Eq.~(\ref{eq:emittance_damp}) can be solved numerically
to obtain the evolution of the vertical
rms emittance during the acceleration.

\begin{figure}[!htb]
   \centering
   \includegraphics*[width=1\columnwidth]{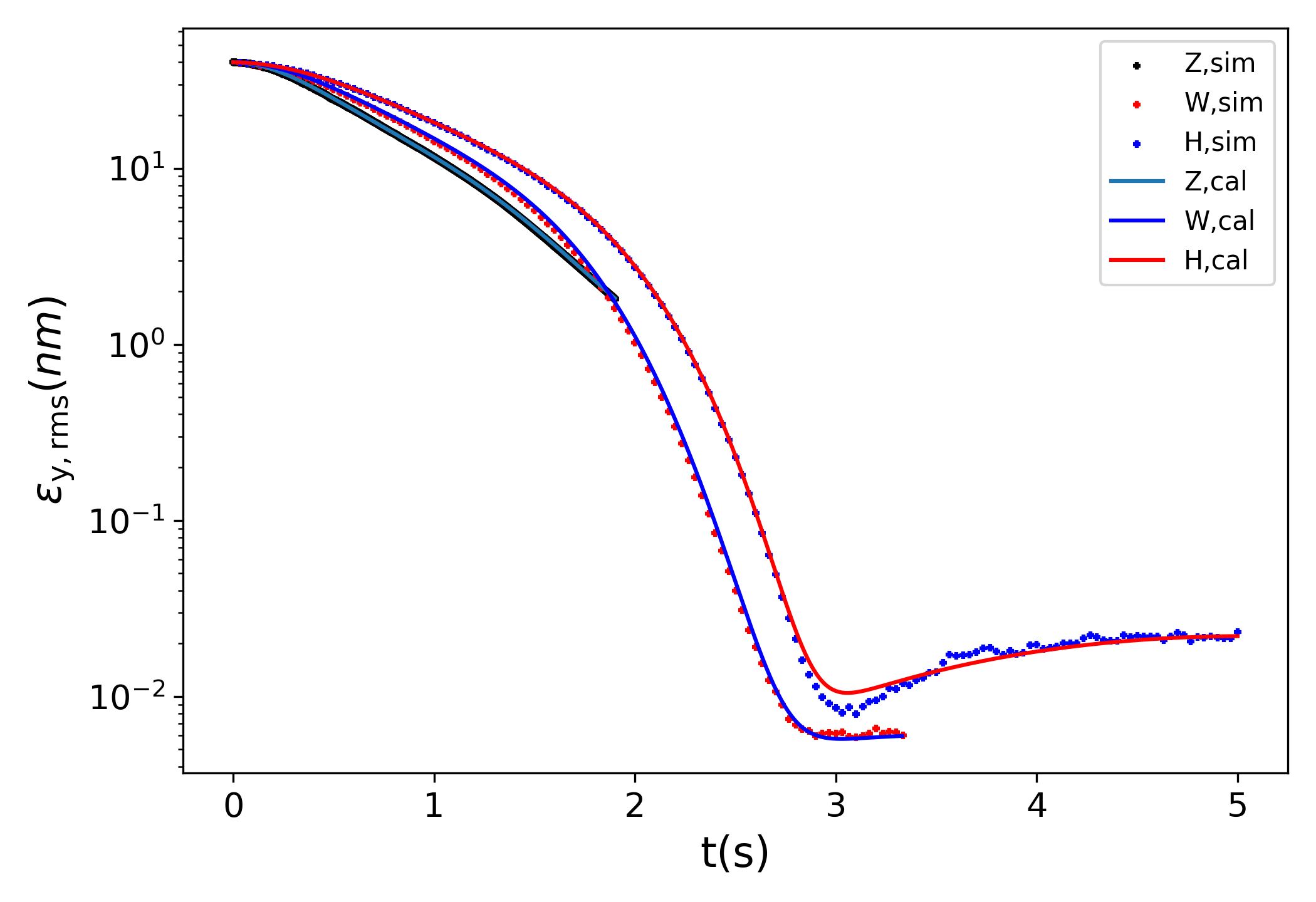}
   \caption{The evolution of the simulated vertical rms
   emittance~(in colored triangle symbols) and the
   calculated vertical rms emittance~(in colored solid
lines) using Eq.~(\ref{eq:emittance_damp}) during
the acceleration for the 3 different operation modes. The error seed 001 in the collection is used,
the vertical equilibrium emittance is 21~pm at 120 GeV.}
    \label{fig:vertical_emittance}
\end{figure}

A more realistic way to evaluate the evolution of the vertical rms emittance
is to carry out multi-particle simulations.
For this we used the ``long\_term\_tracking'' program~\cite{ltt} of Bmad to simulate the
acceleration process. We
set the ramping curve of the beam energy
and the RF parameters, and turned on the
radiation damping and quantum excitation in 
the element-by-element tracking.
We tracked a Gaussian beam with 1000 particles with
the initial parameters specified in Table~\ref{tab:parameters_of_injection_beam}, 
and computed the
eigenemittances along the ramp.
For error seed 001
from the collection, the
evolution of the vertical emittance during the acceleration
is shown in Fig.~\ref{fig:vertical_emittance}.
In the Z-mode, 
the final vertical beam emittance is much larger than the equilibrium vertical emittance at the extraction energy
due to insufficient damping.
In contrast, the final vertical beam emittance
reaches
the equilibrium vertical emittance in the W-mode and H-mode.
The simulation agrees well with the theoretical prediction of Eq.~(\ref{eq:emittance_damp}) for both the Z-mode and W-mode. There is
some discrepancy at around $t=3$~s between the simulation and
the theory for the H-mode, which can be attributed to the
fact that the sawtooth
effect is not included in Eq.~(\ref{eq:emittance_damp}). The above observation also holds for most other error seeds.
In the estimation of depolarization effects described
later in this paper,
we used Eq.~(\ref{eq:emittance_damp}) to obtain the evolution
of vertical rms emittance, as a reasonably good approximation.

\section{STRUCTURE of SPIN RESONANCEs for THE CEPC BOOSTER LATTICE\label{sec:spin_resonance_structure}}

The general lattice structure and the setup of error seeds have already been introduced in Section~\ref{sec:booster_lattice}.
Table~\ref{tab:parameters_of_resonance} lists lattice parameters relevant for analyzing the structure of spin resonances for the CEPC
booster lattice. 

\begin{table}[!htb]
\caption{Key lattice parameters relevant to the structure of spin resonances}
\begin{ruledtabular}
\begin{tabular}{lcdr}

 Vertical tune $\nu_y$ &353.28\\
  Superperiod $P$       &8     \\ 
  Number of FODO cells in each arc section $M$    &140    \\ 
Proportion of total arc bending
angle $\eta_\mathrm{arc}$& 140/142\\
Contribution to $\nu_y$ from all arc sections $\nu_B$&280\\
Number of FODO cells in each straight section $M'$ & 24 \\
Contribution to $\nu_y$ from all straight sections $\nu_{y,\mathrm{str}}$ & 48 \\
Number of FODO cells in each DOM section $M''$ &  6 \\
\end{tabular}
\end{ruledtabular}
\label{tab:parameters_of_resonance}
\end{table}

The locations of super-strong spin resonances can be identified by
using the analysis for the simple model ring in Section~\ref{sec:model_ring}.
Super-strong intrinsic resonances are located at $K=8n\pm 353.28, n \in \mathbb{Z}$ near $K=1136m\pm284, m \in \mathbb{Z}$. As 
these two conditions cannot be met simultaneously, and the enhancement function $\zeta_{M}(\frac{K\eta_\mathrm{arc}\mp\nu_B}{PM})$ varies slowly with $K$ near the
peak, the first super-strong intrinsic resonance is located at
$K=281.28=-9P+\nu_y$,
corresponding to
a beam energy of 123.9~GeV, where the enhancement amplitude
$|E_P^-E_M^-|$ is 924. On the other hand,
super-strong imperfection resonances are generally
those where both conditions
$K=8n\pm 353, n \in \mathbb{Z}$ and $K=1136m\pm284, m \in \mathbb{Z}$
are nearly satisfied. The first super-strong imperfection resonance is located
at $K=281=-9P+[\nu_y]$. Both the first super-strong imperfection and intrinsic
resonances are located beyond the working beam
energy range of the CEPC booster. Nevertheless, the resonances
just below 120~GeV are still near the super-strong resonances and are enhanced. For example,
the enhancement amplitude 
$|E_P^-E_M^-|$ is 127 at the intrinsic resonance
$K=265.28=-11P+\nu_y$, corresponding to
a beam energy of 116.9~GeV. 
In addition,
the strength of intrinsic and imperfection resonances generally increases with the beam energy in the range of 10~GeV to 120~GeV.
 
To verify the above analysis, we
calculated the strengths of
the intrinsic and imperfection resonances
numerically and now discuss their
features. We then analytically estimated the depolarization due to the crossings
of these resonances.

\subsection{Intrinsic resonances}

\begin{figure}[!htb]
   \centering

\subfigure[In a large energy range]{
\begin{minipage}{\columnwidth}
\centering
\includegraphics[width=1\columnwidth]{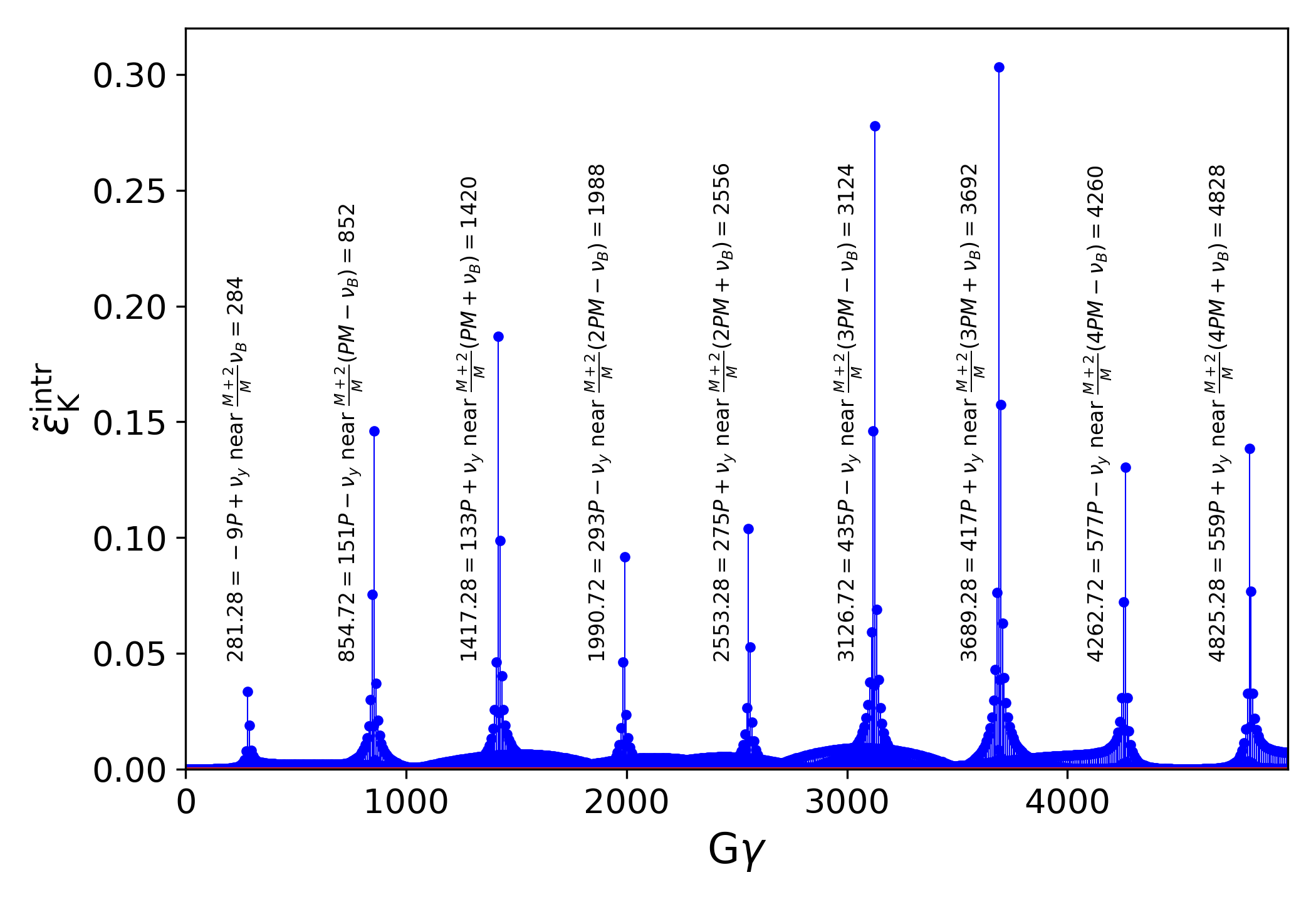} \\
\end{minipage}
}

\subfigure[In the working beam energy range]{
\begin{minipage}{\columnwidth}
\centering
\includegraphics[width=1\columnwidth]{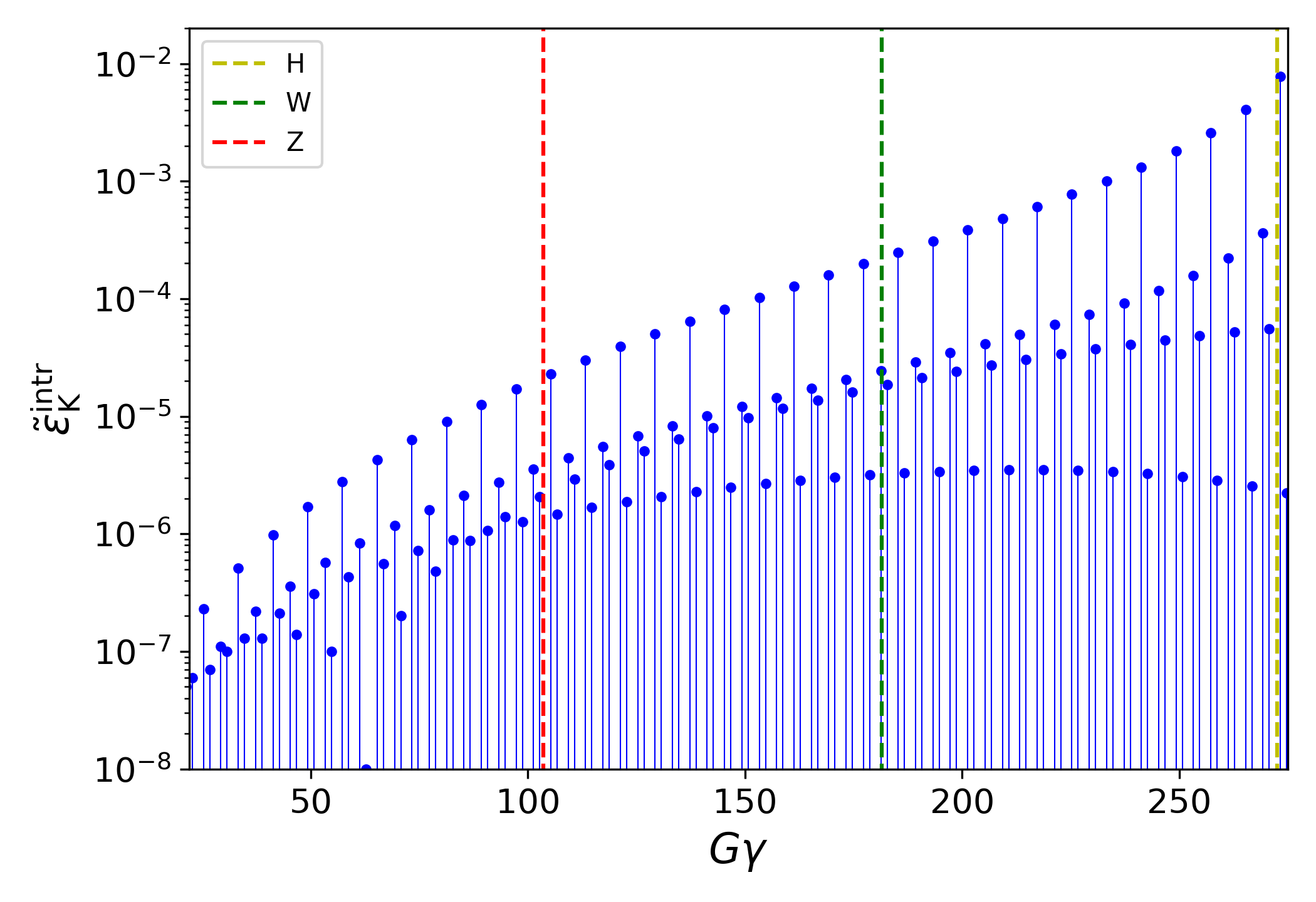} \\
\end{minipage}
}   
  \caption{Intrinsic-resonance spectrum of the bare lattice for a vertical normalized amplitude of $10\pi \textrm{mm} \cdot \textrm{mrad} $. 
  (a) covers a larger energy range
  with the locations of super-strong intrinsic resonances labeled. (b) covers
  the working energy range with the three
  extraction energies marked by dashed lines.
  }
   \label{fig:intrinsic_resonance_strength}
\end{figure}

To calculate the strength of intrinsic resonances
we used the DEPOL code~\cite{courant_acceleration_1980} 
for the bare lattice and the sawtooth effect was not taken into account.
As the strength of an intrinsic resonance
depends on a particle's vertical betatron action
$I_y$, we calculated using a normalized vertical amplitude $\epsilon_{y,\mathrm{norm}}=2\gamma I_y$
of $10 \pi \textrm{mm} \cdot \textrm{mrad}$.
The resulting intrinsic-resonance spectrum is shown in
Fig.~\ref{fig:intrinsic_resonance_strength}.
In Fig.~\ref{fig:intrinsic_resonance_strength}(a), where a large energy range is
covered,
the locations of super-strong intrinsic resonance are labeled by the positions of their peaks. These
agree well with our analysis. Fig.~\ref{fig:intrinsic_resonance_strength}(b)
shows the intrinsic-resonance spectrum in the
working beam energy range. It is clear that
the first super-strong intrinsic resonance is
above 120 GeV, below which the resonance
strength generally increases with energy.
Note that the minor differences in the DOM sections
effectively reduce the periodicity from 8 to 4,
resulting in the adjacent intrinsic resonances spaced by 4 units of $G \gamma$.

As the strength of an intrinsic resonance varies among
particles with different vertical betatron amplitudes
in a beam,
different particles suffer from different levels of 
depolarization after crossing an intrinsic resonance
at the same resonance crossing rate. So, 
to evaluate the beam depolarization,
Eq.~(\ref{eq:F-S_formula}) must be extended
with an ensemble average 
over particles with different vertical betatron amplitudes
for a fixed $\alpha$.
For a Gaussian beam
with an rms vertical emittance of $\epsilon_{y,\mathrm{rms}}$,
the depolarization when crossing a single intrinsic resonance at $\nu_0=K$
is~\cite{leeSpinDynamicsSnakes1997}
\begin{equation}
    \frac{P_f}{P_i}(K, \epsilon_{y,\mathrm{rms}},\alpha)=\frac{1-\frac{\pi {|
    \tilde{\epsilon}_{K}^{\mathrm{intr},\pm}(\epsilon_{y,\mathrm{rms}})
    |}^2}{\alpha}}{1+\frac{\pi {|\tilde{\epsilon}_{K}^{\mathrm{intr},\pm}(\epsilon_{y,\mathrm{rms}})|}^2}{\alpha}}
\end{equation}

Many intrinsic resonances would
be crossed in the acceleration process. For an initial 100\% vertically polarized beam, 
the vertical beam polarization at a certain time $t$ during the
acceleration $P^{\mathrm{intr}}_{\mathrm{trans}}(t)$
can be approximated by multiplying together the surviving polarization
due to each intrinsic resonance the beam encounters before the time $t$,
\begin{equation}
    P^{\mathrm{intr}}_{\mathrm{trans}}(t)\approx\prod_{K \le G\gamma(t)} \frac{P_f}{P_i}(K, \epsilon_{y,\mathrm{rms}},\alpha)
    \label{eq:multiple_intrinsic_depolarization}
\end{equation}
Then, the final polarization after the acceleration is
$P^{\mathrm{intr}}_{\mathrm{trans},f}=P^{\mathrm{intr}}_{\mathrm{trans}}(t_{\mathrm{ramp}})$.

\begin{figure}[tbh!]

   \centering

\subfigure[Spectrum of scaled intrinsic resonances]{
\begin{minipage}{\columnwidth}
\centering
\includegraphics[width=1\columnwidth]{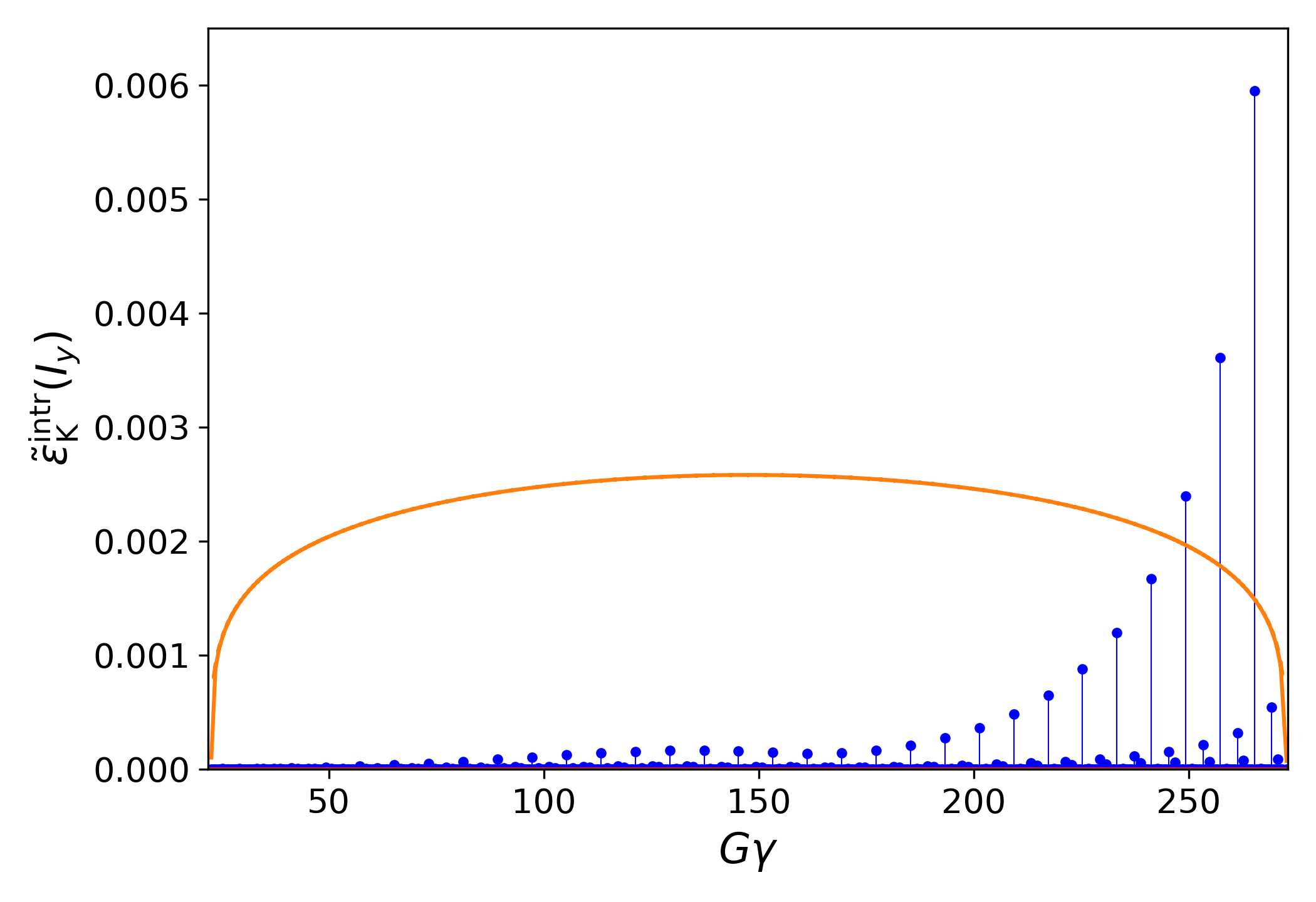} \\
\end{minipage}
}
\subfigure[Depolarization due to intrinsic resonances]{
\begin{minipage}{\columnwidth}
\centering
\includegraphics[width=1\columnwidth]{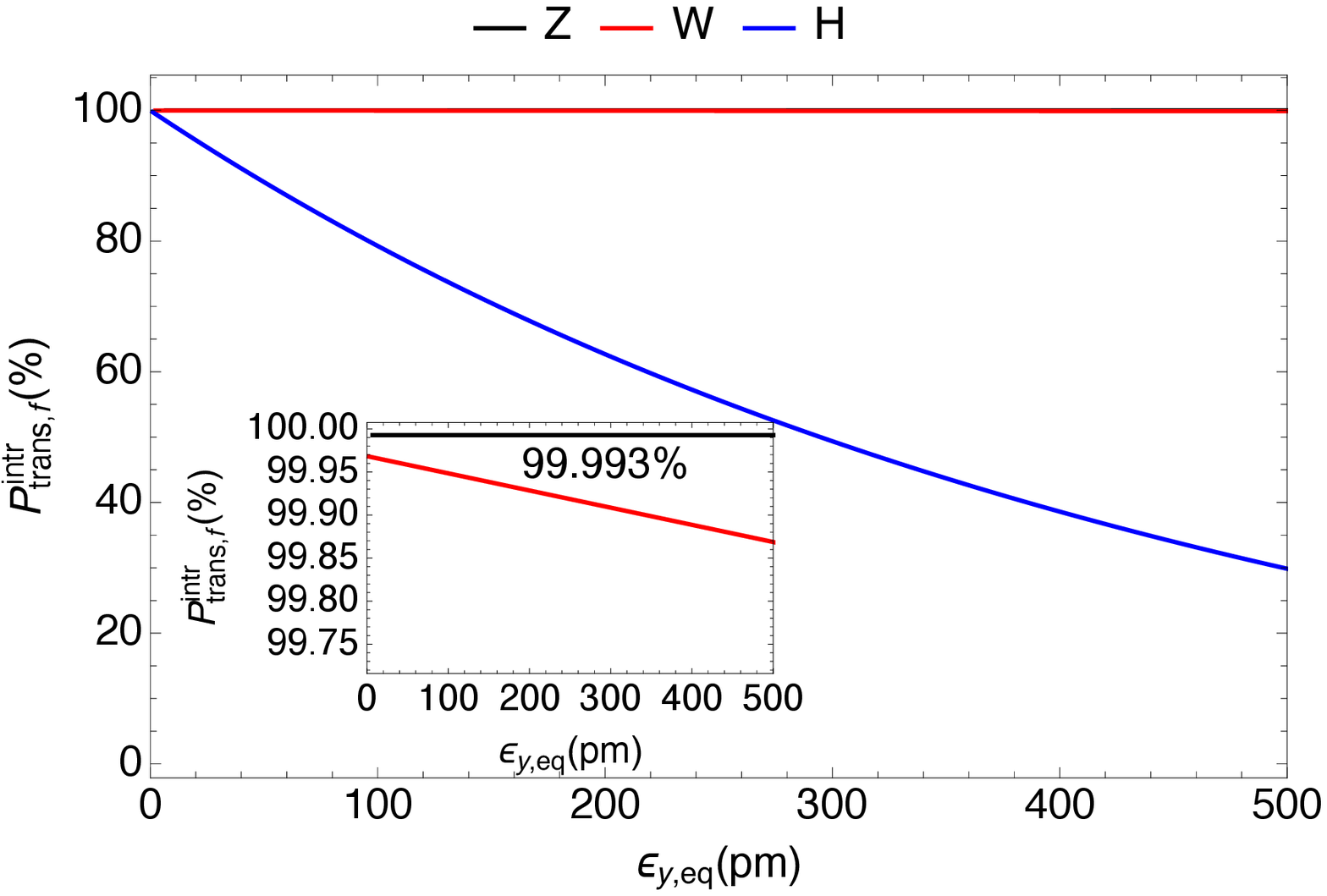} \\
\end{minipage}
}   
   \caption{ (a) shows the
   intrinsic-resonance spectrum in the working beam energy
   range, scaled using the vertical rms emittance at the time of resonance crossing,
   assuming a vertical equilibrium emittance of 100~pm at 120 GeV. The orange line 
   depicts the estimated
   upper strength limit for each intrinsic resonance needed to ensure the depolarization due to its crossing of less than 1\%.
   (b) shows the estimated final vertical polarization after crossing all the intrinsic resonances during the acceleration process, for different vertical equilibrium emittances $\epsilon_{y,\mathrm{eq}}$ at 120 GeV, and 
   for the three operation modes. The inset plot shows a zoom-in
   view of the results of the Z-mode and the W-mode,
   with tiny depolarization. In contrast, the H-mode
   suffers from more severe depolarization in particular for a larger $\epsilon_{y,\mathrm{eq}}$.}
   \label{fig:emit_vs_P_trans_int}
\end{figure}

As shown in Fig.~\ref{fig:seed_emity}, the vertical equilibrium emittance
at a beam energy of 120~GeV $\epsilon_{y,\mathrm{eq},0}$
varies among different error seeds, in the range of 6~pm to about 500~pm.
The vertical rms emittance also varies during the acceleration process as shown in Fig.~\ref{fig:vertical_emittance}.
So we chose a vertical equilibrium emittance of 100~pm at 120 GeV, and evaluated
the evolution of the vertical rms emittance during the acceleration process
according to Eq.~(\ref{eq:emittance_damp}). For
each intrinsic resonance in the working beam energy range,
we scaled its strength according to the vertical rms emittance at the time of crossing, as shown in Fig.~\ref{fig:emit_vs_P_trans_int}(a). Using Eq.~(\ref{eq:F-S_formula}) and the spin-resonance crossing rate according to Eq.~(\ref{eq:energy_ramp}), we obtained the upper strength limit for
each intrinsic resonance needed to ensure the depolarization due to its crossing of less
than 1\%. This is represented
as the orange curve in Fig.~\ref{fig:emit_vs_P_trans_int}(a). When compared with the spectrum of scaled intrinsic resonances, it is clear 
that only three intrinsic resonances near 120 GeV can cause severe depolarization.

Next,
we scanned a range of vertical equilibrium emittance
$\epsilon_{y,\mathrm{eq},0}$
at a beam energy of 120~GeV. For
each case, we used Eq.~(\ref{eq:emittance_damp}) to evaluate the vertical rms emittance evolution
during the acceleration.
Then we used Eq.~(\ref{eq:multiple_intrinsic_depolarization}) to estimate the polarization loss after crossing
all intrinsic resonances in the acceleration process.
The spin-resonance crossing rate was computed with Eq.~(\ref{eq:energy_ramp}).
The results for the three operation modes are shown in Fig.~\ref{fig:emit_vs_P_trans_int}(b).
In the Z-mode, the vertical rms emittance remains far from reaching
the vertical equilibrium emittance during the short ramping time, and
the change in the evolution of the beam vertical rms emittance is small
for the range of $\epsilon_{y,\mathrm{eq}}$ we considered. In addition,
the strengths of intrinsic resonances are generally small,
resulting in a polarization loss less than 0.01\% and a slow variation as a function of $\epsilon_{y,\mathrm{eq}}$.
However, in the W-mode and H-mode, the vertical rms emittance 
reaches $\epsilon_{y,\mathrm{eq}}$ during acceleration and the polarization loss increases with $\epsilon_{y,\mathrm{eq}}$.
The polarization loss in the W-mode is still less than 0.15\%,
while the polarization loss in the H-mode 
depends strongly on $\epsilon_{y,\mathrm{eq}}$, and 
exceeds 20\% for $\epsilon_{y,\mathrm{eq}}$ larger than 100~\si{pm}, mainly due to the crossings of several
strong intrinsic resonances before 120~GeV. 
This highlights the importance of reducing $\epsilon_{y,\mathrm{eq}}$ to small values through optics
correction, to minimize the depolarization caused by intrinsic resonance crossings.

\subsection{Imperfection resonances}

To calculate the strengths of imperfection resonances for 
different error seeds,
We implemented the algorithm of the DEPOL code~\cite{courant_acceleration_1980} in
a Mathematica script~\cite{Mathematica} 
using the closed orbit and optics functions calculated by Bmad, where
the sawtooth effect was not taken into account. The calculation took into account the major contributions due
to quadrupoles, bends and sextupoles, as well as corrector fields inside the quadrupoles,
but didn't consider the magnet roll errors nor the dipole relative field errors.

\begin{figure}[!htb]
   \centering

\subfigure[In a large energy range]{
\begin{minipage}{\columnwidth}
\centering
\includegraphics[width=0.95\columnwidth]{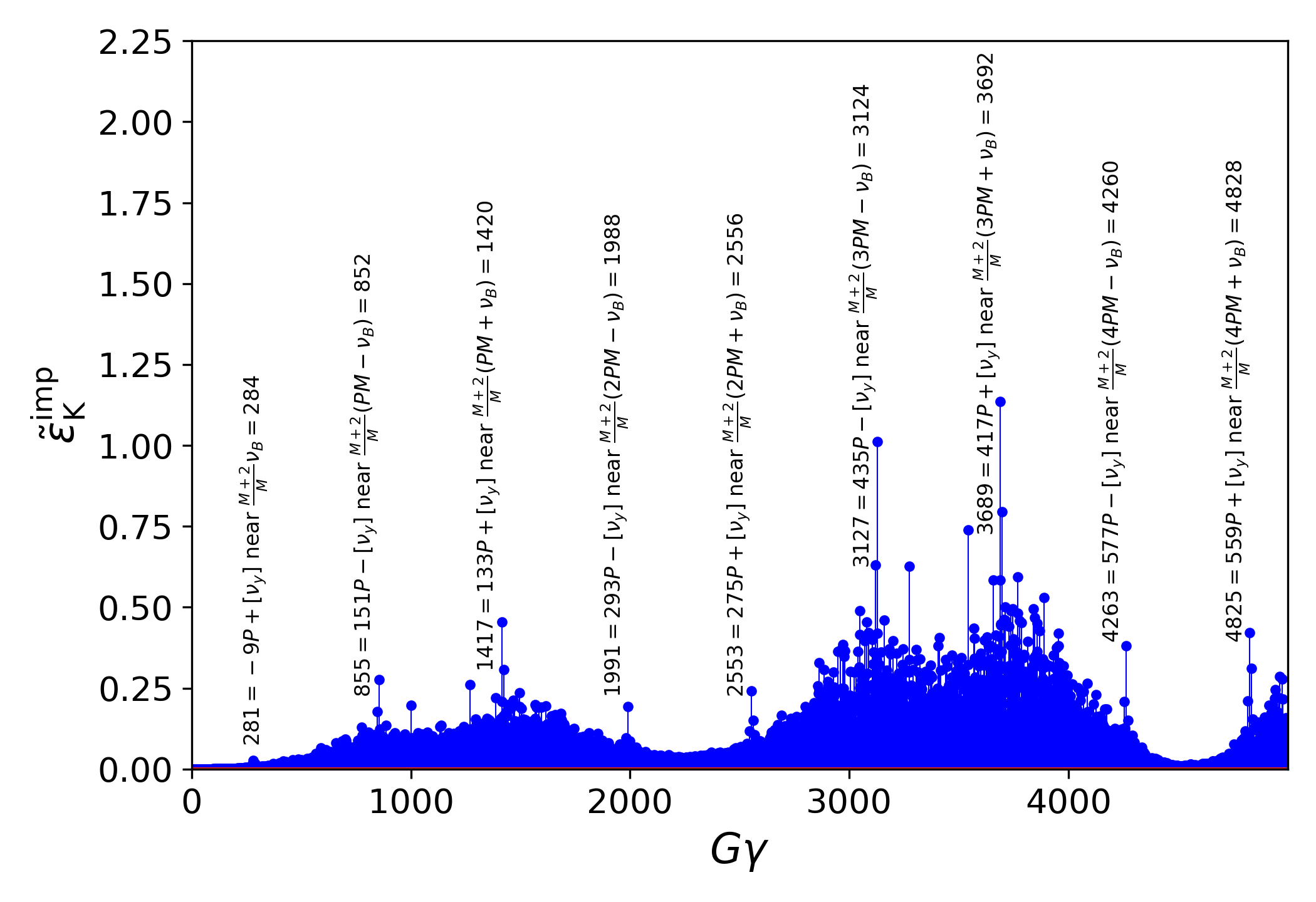} \\
\end{minipage}
}

\subfigure[In the working beam energy range]{
\begin{minipage}{\columnwidth}
\centering
\includegraphics[width=0.9\columnwidth]{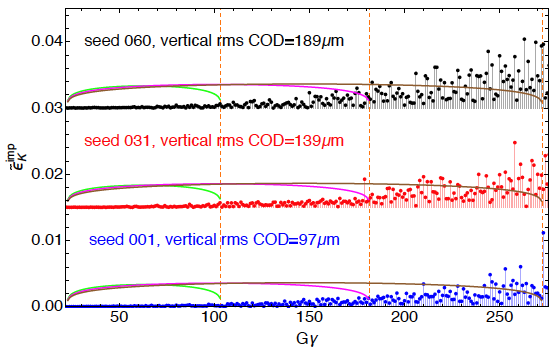} \\
\end{minipage}
}   
  \caption{Imperfection-resonance spectra of several error seeds
  of the CEPC booster lattice.
  (a) shows the imperfection-resonance spectrum of error seed 001 in a large energy range, highlighting the locations of super-strong imperfection resonances.
  (b) shows the imperfection-resonance spectra of error seed 001, 031 and 060 with increasing vertical rms CODs in the working energy range, an offset of 0.015
  is appended to the latter two cases for comparison. The three dashed
  vertical lines denote the three extraction energies.
  The green, magenta
  and brown curves depict the upper strength limit for
  each imperfection resonance needed to ensure the depolarization
  due to its crossing of less than 1\%, for the Z-mode, W-mode, H-mode, respectively.}
   \label{fig:imperfection_resonance}
\end{figure}

Fig.~\ref{fig:imperfection_resonance}(a) shows the imperfection-resonance spectrum in a large energy range for error seed 001 with a vertical
rms COD of 97~\si{\um}. The locations of the super-strong imperfection
resonances mostly agree with our previous analysis, apart from a few peaks corresponding to 
$|k|\neq [\nu_y]$.
Since the strength of an imperfection
resonance is a superposition of various Fourier terms of the
vertical COD, there is, in addition, a broad ``background'' near each of these super-strong resonances. 
Fig.~\ref{fig:imperfection_resonance}(b)
illustrates the imperfection-resonance spectra for three error seeds with different vertical rms CODs in the working energy range. The
strength of imperfection resonances generally increase with energy, with 
120 GeV just below the first strong imperfection resonance. 
Besides,
it is clear there is an increase in
the strength of the strongest imperfection resonance just below 120 GeV for error seed 060 with the lagest vertical rms COD.

Unlike intrinsic resonances, the strength of an imperfection resonance is the same for all the particles in a beam, so that the polarization loss due to crossing an isolated imperfection resonance can be estimated directly
using Eq.~(\ref{eq:F-S_formula}).  
For each imperfection resonance, we calculated 
the upper strength limit for which the
depolarization due to its crossing is less than 1\%,
considering the resonance crossing rate
as shown in Fig.~\ref{fig:ramping_curve}.
We then obtained three curves
of the upper strength limit for each imperfection resonance,
colored in green, magenta and brown for the Z-mode, W-mode and H-mode in Fig.~\ref{fig:imperfection_resonance}, respectively.
These curves can be compared with the strengths of imperfection resonances to
identify the most dangerous resonances whose strengths
lie above these curves.
Clearly, the imperfection resonances near the extraction energies 
have larger strengths, but are crossed at a lower speed, potentially
causing more severe depolarization.
For the Z-mode, all
imperfection resonances in the three error seeds are below the green curve.
For the W-mode, the strengths of only a few imperfection resonances
in error seeds 031 and 060 lie above the magenta curve, while 
the strengths of all imperfection resonances in error seed 001
are below the magenta curve. In contrast, for the H-mode, 
the strengths of a significant number of imperfection resonances lie above the brown curve for the three error seeds, potentially causing
substantial depolarization.

We quantify the depolarization due to the crossings of imperfection
resonances, using a method similar to that for intrinsic resonances.
For an initial 100\% vertical polarization, 
we can estimate
the vertical polarization at a specified time $t$ during
acceleration, $P^{\mathrm{imp}}_{\mathrm{trans}}(t)$,
by multiplying together the surviving polarization due
to each imperfection resonance that the beam encounters
before the time $t$,
\begin{eqnarray}
    P^{\mathrm{imp}}_{\mathrm{trans}}(t)&\approx&\prod_{K \le G\gamma(t)} \frac{P_f}{P_i}(K, \alpha) \nonumber \\
    &=&\prod_{K \le G\gamma(t)}
     \left[2\exp(-\frac{\pi \vert {\tilde \epsilon}_K^{\textrm{imp}}  \vert^2}{2\alpha})-1\right]
    \label{eq:multiple_resonance_loss_imp}
\end{eqnarray}
Then, the final vertical polarization after the acceleration for a time $t_{\mathrm{ramp}}$ is
$P^{\mathrm{imp}}_{\mathrm{trans},f}=P^{\mathrm{imp}}_{\mathrm{trans}}(t_{\mathrm{ramp}})$. This implicitly assumes there
is no correlation in the depolarization
due to successive crossings
of adjacent imperfection resonances.

\begin{figure}[!htb]
    \includegraphics*[width=1\columnwidth]{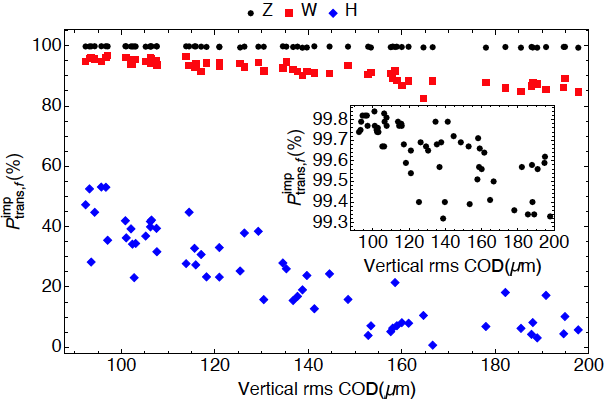}
   \caption{Estimated final vertical beam polarization after crossing all the imperfection resonances during the acceleration process
   in the three operation modes, for
   the 60 error seeds with different vertical rms CODs. The inset plot shows a zoom-in view of the result for the Z-mode, with tiny depolarization. In contrast, the depolarization is more
   severe in the W-mode and the H-mode, especially for the error seeds with larger vertical rms CODs.}
   \label{fig:orbit_vs_P_trans_imp}
\end{figure}

We calculated the strengths of imperfection resonances for all 60 error seeds, and estimated the depolarization due to the
crossings of these imperfection resonances, for the three operation modes. The results,
shown in Fig.~\ref{fig:orbit_vs_P_trans_imp}, 
indicate that the polarization loss generally increases with the vertical rms COD.
However, the polarization loss also varies among error seeds with similar vertical rms CODs,
due to the difference in the spectra and
maximum strengths of the imperfection resonances.
For the Z-mode and the W-mode, the depolarization is less than 1\% 
and 20\%, respectively. In contrast,
in the acceleration to 120 GeV, the depolarization is
much more severe, ranging from about 50\% to almost complete depolarization.
This is due to the presence of many dangerous imperfection resonances near 120~GeV, as shown in Fig.~\ref{fig:imperfection_resonance}.

In addition, the polarization loss
due to these imperfection resonances is generally more severe than that caused by intrinsic resonances as estimated in the previous subsection.
The next section will present an analysis of the combined effects of 
both imperfection and intrinsic resonances, as well as more realistic multi-particle simulation results of the beam depolarization in the acceleration process.

\section{EVALUATION OF THE POLARIZATION TRANSMISSION\label{sec:simulations}}
In this section
we present quantitative evaluations of the beam polarization
transmission in the CEPC booster. Assuming the injected
beam has a 100\% vertical polarization, 
the evolution of the vertical beam polarization in the acceleration
process, $P_{\mathrm{trans}}(t)$, and
the final vertical beam polarization, $P_{\mathrm{trans},f}$,
are evaluated using two different approaches.

In the previous section, we estimated the depolarization
during acceleration in the CEPC booster,
by separately considering
the contributions of intrinsic resonances and imperfection resonances.
By combining these two contributions, we can now estimate the overall 
depolarization effects due to both types of resonances,
\begin{equation}
P_{\mathrm{trans}}(t)\approx P^{\mathrm{intr}}_{\mathrm{trans}}(t)\times P^{\mathrm{imp}}_{\mathrm{trans}}(t)
\end{equation}
and the final beam polarization after the acceleration is
$P_{\mathrm{trans},f}=P_{\mathrm{trans}}(t_{\mathrm{ramp}})$.
We applied such estimation for all the 60 error seeds and we used Eq.~(\ref{eq:energy_ramp}) to evaluate
the evolution of beam energy and resonance crossing rate.
For each error seed, we computed its
vertical equilibrium 
emittance at a beam energy of 120~GeV $\epsilon_{y,\mathrm{eq},0}$,
and then evaluated the evolution of the vertical
rms emittance $\epsilon_{y,\mathrm{rms}}$ using Eq.~(\ref{eq:emittance_damp}).
Note that the strengths of
intrinsic resonances 
of these error seeds were approximated by those of
the bare lattice
as shown in Fig.~\ref{fig:intrinsic_resonance_strength}.
We scaled the strengths of intrinsic resonances during the acceleration
process taking into account the evolution of $\epsilon_{y,\mathrm{rms}}$.
In the following context, we refer to the $P_{\mathrm{trans},f}$ obtained 
using this method as the estimation results.

\begin{figure}[!htb]
   \centering
   \includegraphics*[width=1\columnwidth]{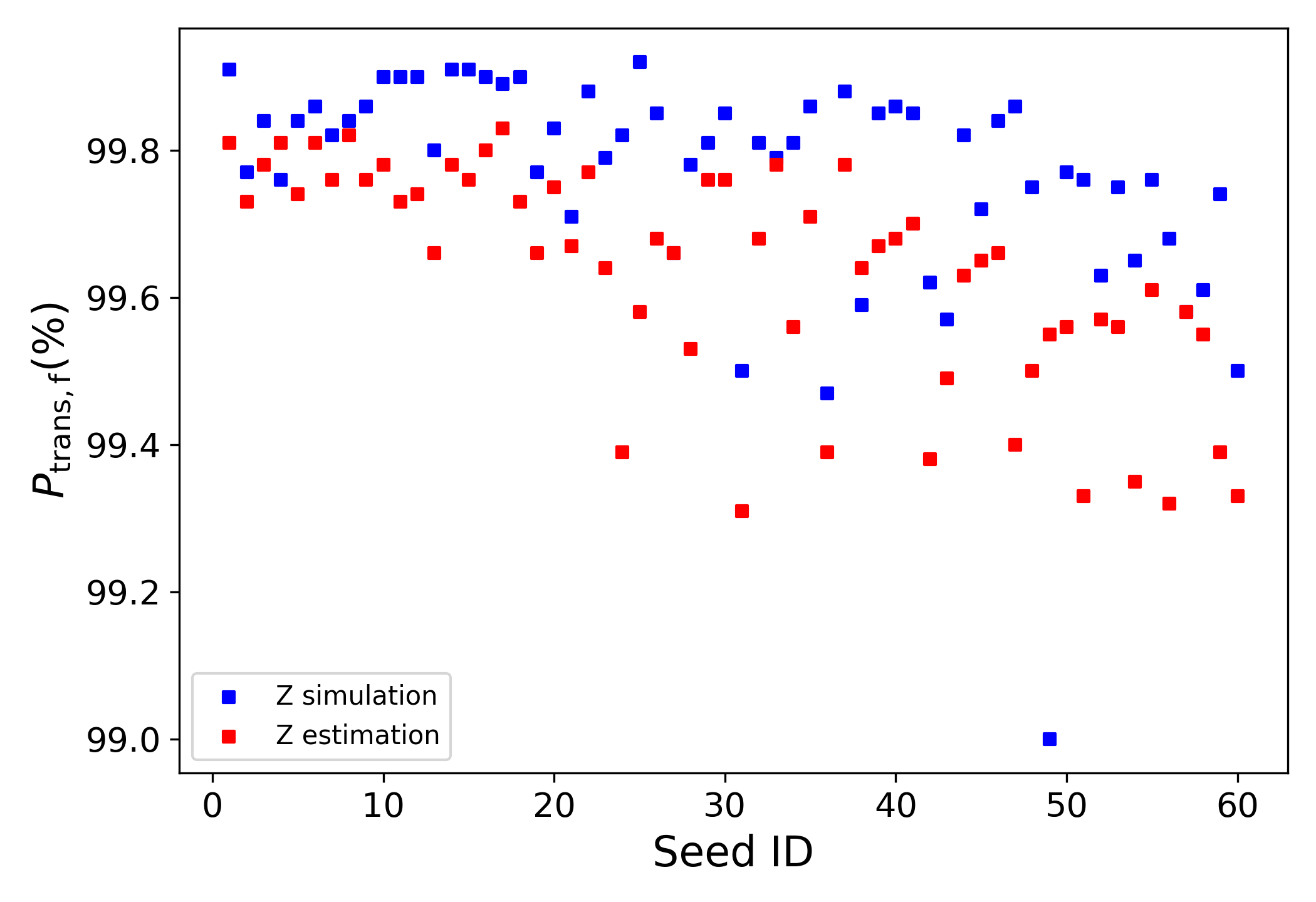}
      \includegraphics*[width=1\columnwidth]{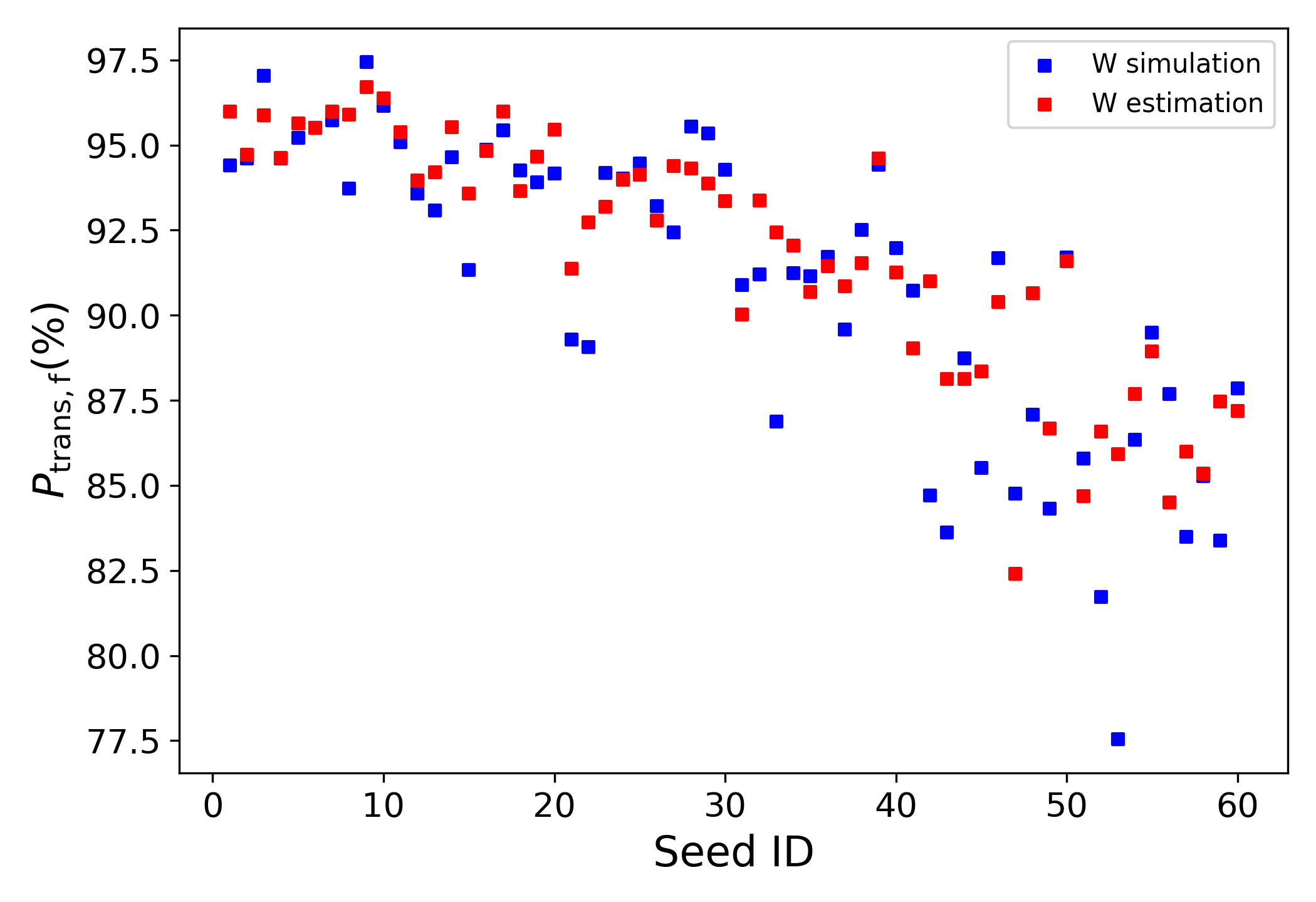}
         \includegraphics*[width=1\columnwidth]{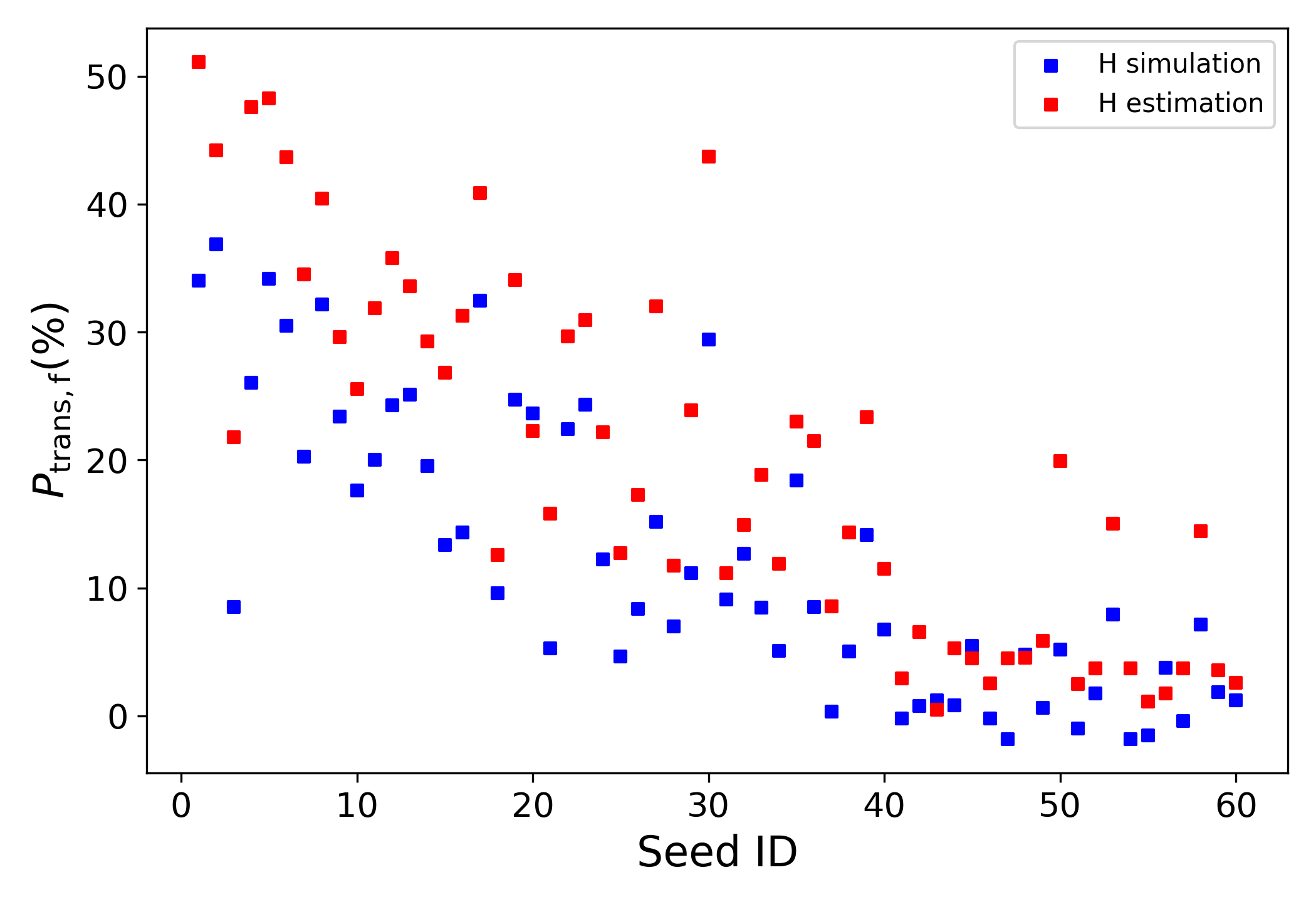}
   \caption{The final beam polarization
   after the acceleration $P_{\mathrm{trans},f}$
   for the 60 error seeds, assuming the injected beam is 100\% vertically polarized.
   The top, middle and bottom plots are for the Z-mode, the W-mode and
   the H-mode, respectively.
   The blue and
   red points represent the results of simulation and estimation,
   respectively.}
   \label{fig:depolarization_simulation_calculation}
\end{figure}

To verify our analysis, we utilized 
more realistic multi-particle tracking simulations
of the acceleration process.
For this we used the ``long\_term\_tracking'' option of BMAD and we tracked a Gaussian beam of 1000 particles with an initial vertical polarization of 100\%. 
We set the ramping
curves of the beam energy and the RF parameters, and
turned on the radiation damping and quantum excitation effects in the element-by-element tracking with
full six-dimensional orbit motion and three-dimensional spin motion. Hereafter we refer to the $P_{\mathrm{trans},f}$ obtained using this method as the simulation results.

The estimates and simulation results of $P_{\mathrm{trans},f}$ for
the 60 error seeds and the three
operation modes are shown in Fig.~\ref{fig:depolarization_simulation_calculation}. The simulation results for the Z-mode and
the W-mode show maintenance of a high level of beam polarization, with $P_{\mathrm{trans},f}$ 
mostly above 99\% and 80\%, respectively. In contrast, the
simulation results for the H-mode
indicate substantial depolarization at higher beam energies, with
$P_{\mathrm{trans},f}$ generally below 50\%.
The trend observed is that depolarization becomes more severe for larger
seed IDs, 
whose vertical rms CODs are generally larger and imperfection resonances
are generally stronger. 
The estimates of
$P_{\mathrm{trans},f}$
are quite close to the simulation results, for the Z-mode and the W-mode, while the difference
becomes more substantial for the H-mode. It is worth noting that the estimation
method used here, by simply multiplying together
the surviving polarization due to each imperfection and intrinsic resonance, already provides a reasonably accurate order-of-magnitude
estimate of $P_{\mathrm{trans},f}$. This suggests that single crossings of the imperfection
and intrinsic resonances are the major factors of
depolarization in CEPC booster.

\begin{figure}[!htb]
   \centering
   \includegraphics*[width=1\columnwidth]{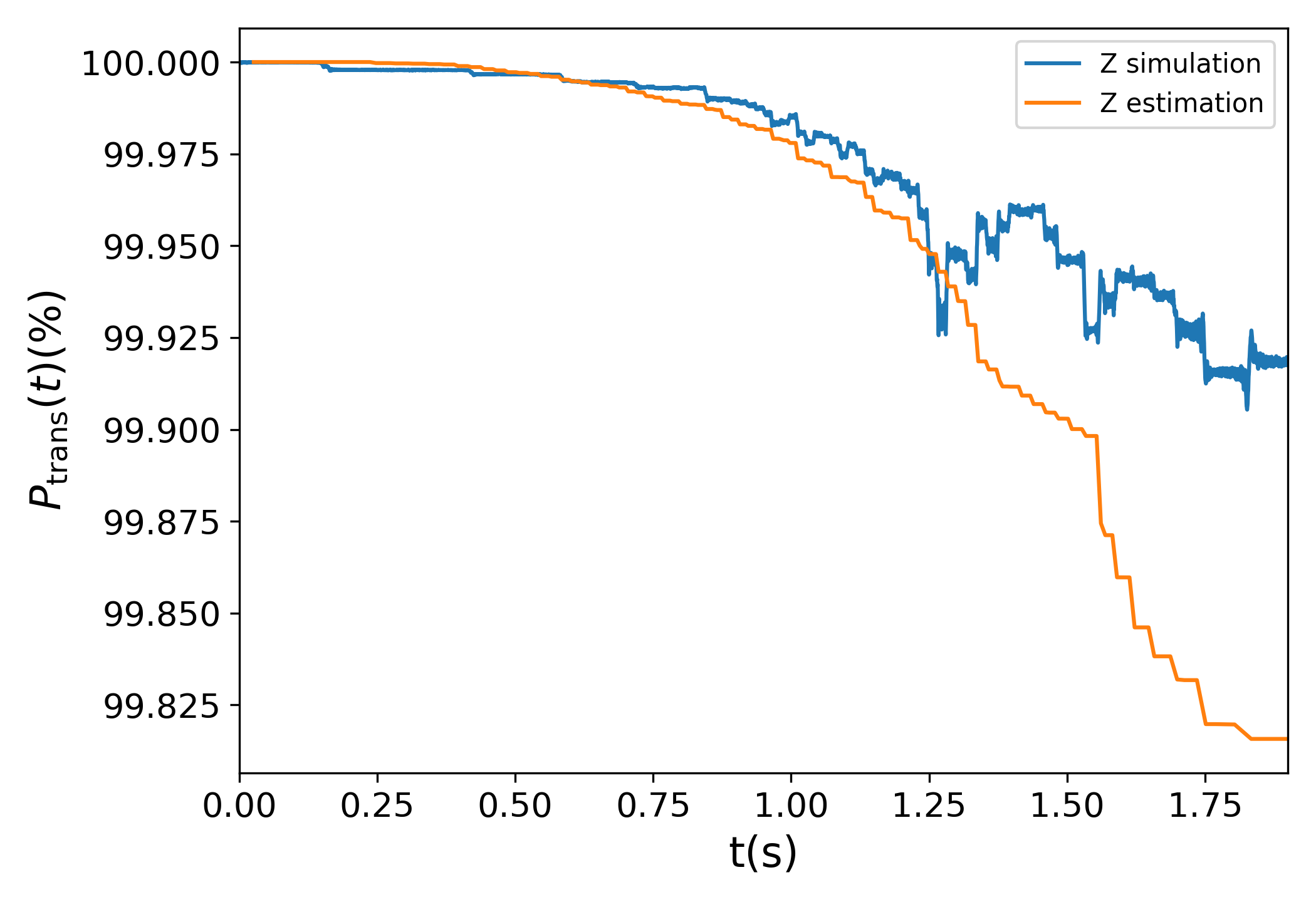}
   \includegraphics*[width=1\columnwidth]{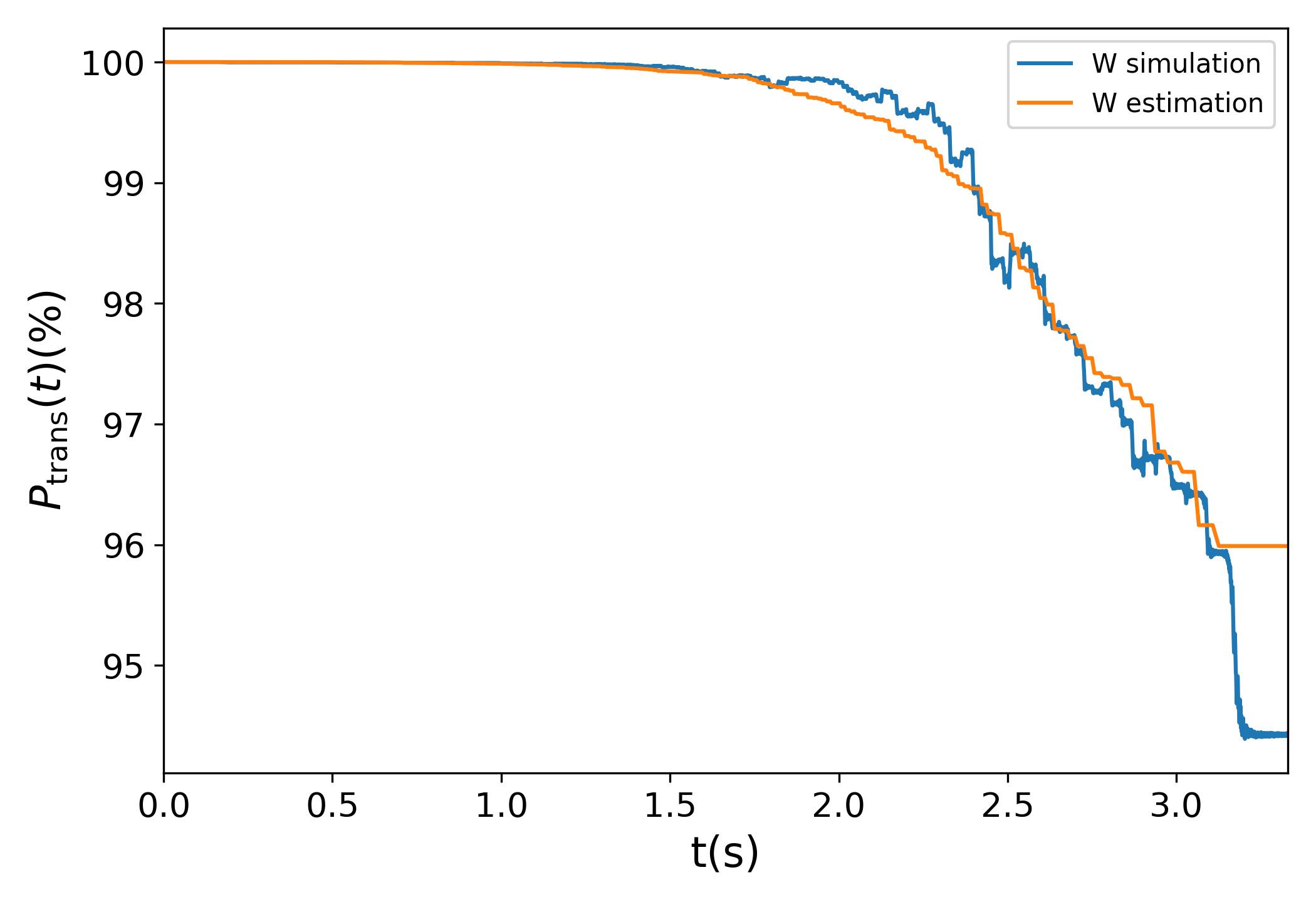}
   \includegraphics*[width=1\columnwidth]{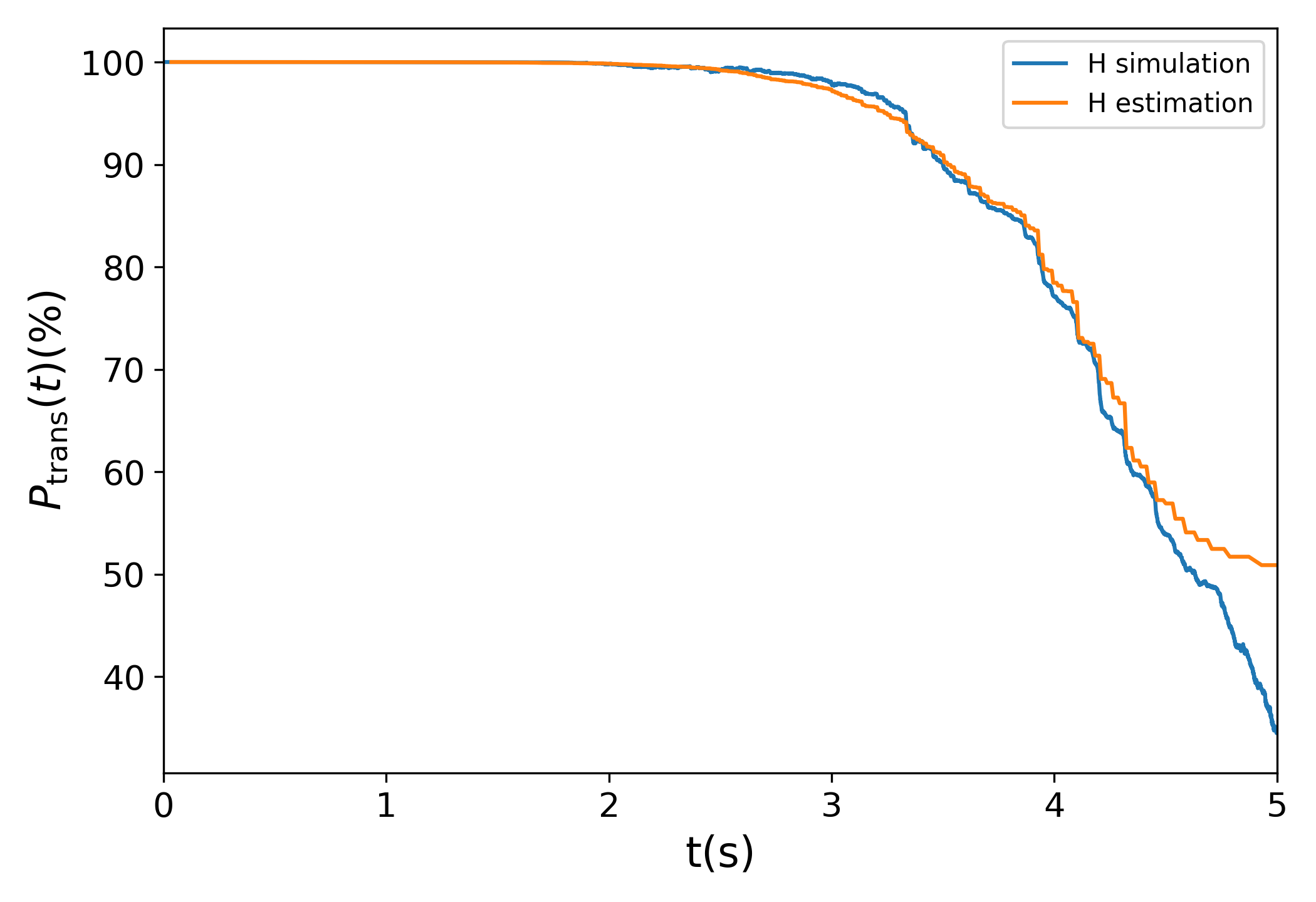}

   \caption{Comparison between the simulated and estimated evolution of $P_{\mathrm{trans}}(t)$ for error
   seed 001 in the Z-mode (top plot), 
   the W-mode (middle plot) and the H-mode (bottom plot), respectively.}
   \label{fig:Sy_evolution}
\end{figure}

To better appreciate the depolarization effects,
the evolution of $P_{\mathrm{trans}}(t)$ in the acceleration process
for a typical error seed 001 is illustrated
in Fig.~\ref{fig:Sy_evolution}. Most polarization loss occurs
near the end of the acceleration process, when the ramping rate decreases while the resonance strengths
become larger as the beam energy increases. Although the estimation results
and simulation results of the $P_{\mathrm{trans},f}$ are generally
similar, there are noticeable discrepancies in the evolution of 
$P_{\mathrm{trans}}(t)$. Next, we discuss several possible sources
of the discrepancies.

Firstly, the method of calculation
we adopted for the strengths of spin resonances
has limitations. 
For the calculation of the strength of imperfection resonances,
the sawtooth orbit was not taken into account,
nor were the magnet roll errors or the dipole relative field errors, which could lead to some
discrepancies and affect the estimation of depolarization. Moreover, 
we used the strengths of intrinsic resonances for the
bare lattice in the estimation of $P_{\mathrm{trans}}(t)$ for
error seeds.
The depolarization from intrinsic
resonances is very weak for the Z-mode and W-mode, and
the difference in the strength of intrinsic resonances does not
affect the result much. However,
the discrepancies
between the bare lattice and the imperfect lattice could impact the results for the H-mode.  
We are aware that the strengths of first-order spin resonances of error
seeds can be more precisely
calculated~\cite{hoffstaetter_high-energy_2006} within Bmad~\cite{Bmad}, this approach will be used in the future study of the depolarization effects in the H-mode.

Second, the simulation results exhibit
fluctuations in $P_{\mathrm{trans}}(t)$,
most obvious in the Z-mode, 
which are a signature of
the interference-overlap effect~\cite{chao_spin_2007}
between the successive crossings of multiple adjacent imperfection resonances. 
This effect seems to be smeared out at higher beam energies,
due to the loss of coherence of the spin phase.
One potential contributor is the stochastic change of particle
energy due to synchrotron radiation, which was implemented on an element-by-element basis in the tracking with Bmad.
In contrast, 
this effect was not taken into account
in the estimates of depolarization,
resulting in monotonic decrease
of $P_{\mathrm{trans}}(t)$ in the acceleration process.

Third, in our estimates, we only considered 
the depolarization due to single crossings of
imperfection and intrinsic resonances. In contrast, the simulation results
also take into account
various more complicated depolarization effects, such as
higher-order spin resonances, multiple crossings of the same
underlying spin resonances due to synchrotron oscillations~\cite{aniel_polarized_1985,yokoya_effects_1983},
as well as radiative depolarization effects~\cite{dks,yokoya_effects_1983,xia_evaluation_2022}, etc.
If we compare, for the Z-mode and the W-mode,
the discrepancy between the estimates and simulation results, 
assuming they are mostly due to these alternative sources, then according to
Fig.~\ref{fig:depolarization_simulation_calculation}(a) and (b), 
we can estimate that their contribution to
depolarization is no more than
a few percent.
However, it is more difficult to estimate their contribution for the H-mode, as the
discrepancy shown in Fig.~\ref{fig:depolarization_simulation_calculation}(c) is much larger. Better understanding of these effects at
higher energies near 120 GeV is needed and will be addressed elsewhere.

In addition, there is also a concern that the 
fluctuation of magnetic fields due to dynamic effects
during the ramping process could cause additional
depolarization. Dipole magnets, whose dynamic errors mostly affect the horizontal beam orbit, are typically powered in series, further reducing their influence. Dynamic errors of
correctors and quadrupoles are more of concern. To this end,
we added additional
random field fluctuations to correctors and 
quadrupoles of up to 0.2\%, to five error seeds from the 
collection with vertical rms CODs ranging from 100\si{\um} to 180\si{\um},
and evaluated the influence
to the polarization transmission for the Z-mode and W-mode.
For each error seed and each amplitude of random field
fluctuation, we generated 10 different random seeds of 
additional field errors.
The variation in the polarization transmission is
tiny for the Z-mode, and less than 3\% for the W-mode.
Though random field fluctuations were implemented as fixed
values rather than more realistic
dynamic variations during the ramping process,
these simulations show the polarization transmission is
not sensitive to typical levels
of additional field fluctuations, for the Z-mode and W-mode.

\section{CONCLUSION}

This paper examines the depolarization effects
during the acceleration process in the CEPC booster,
from an injection energy of 10 GeV,
to the extraction energies of 45.6~GeV, 80~GeV
and 120~GeV, in the three operation modes,
respectively. We used a simplified 
lattice for the CEPC booster, introduced machine
imperfections and launched 
closed-orbit correction and betatron-tune
correction, generating 60 error
seeds for evaluation of depolarization effects. We studied
the structure
of the imperfection and intrinsic spin
resonances for a simplified
lattice model analytically, and then applied the analysis to
the CEPC booster lattice.
The locations of super-strong resonances and
the general behavior of the resonance
strengths in the working beam energy
range were then verified with numerical
calculations of the strengths of spin resonances
for these 60 error seeds.
The depolarization due to the crossings
of these imperfection and
intrinsic resonances were estimated
using the Froissart-Stora formula, and compared with the results of 
multi-particle tracking simulations
of the acceleration process. These studies
suggests that a high level of beam polarization can be
maintained during the acceleration to 45.6~GeV and 80~GeV,
but severe depolarization could occur at higher beam energies in the acceleration to 120~GeV.

Our main finding is
that the CEPC booster lattice has a high
``effective''
periodicity in terms of the lattice contributions to the
strength of imperfection and intrinsic resonances,
similar to that of the EIC booster~\cite{Ranjbar_2018}.
Among imperfection and intrinsic resonances,
super-strong resonances, where the contributions of
all arc FODO cells add up coherently,
occur near 
$(mPM\pm \nu_B)/\eta_{\mathrm{arc}},
m \in \mathbb{Z}$.
The spacings between adjacent super-strong resonances
are very large, while the first super-strong resonances
near $\nu_B/\eta_{\mathrm{arc}}$ also correspond to 
very high beam energies.
Meanwhile,
the contributions from a large number of
FODO cells mostly cancel out for the resonances
at much lower beam energies, away from the super-strong resonances.
The key feature
of the CEPC booster lattice is that all super-strong spin resonances are beyond the working
beam energy range, though a few spin resonances near
120 GeV are still much enhanced. 
In fact, our test with several other candidate lattices of CEPC booster and collider rings
confirmed that such a structure of spin resonances is a general feature of future 100~km-scale electron rings. 
This feature can be
further exploited in the lattice design and optimization of these electron rings.

Apart from these underlying structure of spin resonances, the strengths of the imperfection
resonances in the working beam energy range
depend on the amplitude of radial magnetic fields around the ring,
most importantly due to the vertical orbit offsets in quadrupoles.
In this paper, we focus
on the random misalignment errors of
magnets, and implemented
closed-orbit correction to set a 
reasonable range of potential levels of COD, realized in past and
existing machines, like the LEP~\cite{dehning_beam_2004}.
Nevertheless, the
systematic misalignment errors,
due to variations in the vertical magnet positions after the
smoothing procedure of alignment, or uneven settling of
the accelerator floor over time,
could also contribute to the radial magnetic
fields, and affect the strengths of imperfection resonances.
This aspect requires a dedicated
study and is beyond the scope of this paper. 
In addition, we are aware that dedicated orbit bumps can be set up to partially compensate certain strong imperfection resonances, namely the harmonic closed orbit spin matching~\cite{ rossmanithCompensationDepolarizingEffects1985,khiari_acceleration_1989,barberHighSpinPolarization1994}, which shall also be pursued.

The sawtooth effect introduces nontrivial perturbations to the
closed orbit and optics in the CEPC booster at higher beam energies.
Its influence was taken into account in the simulation results of polarization transmission,
but not in the calculation of
strengths of imperfection and intrinsic resonances, and thus not in
the estimation results.
As the discrepancies between the simulation results and estimation
results suggest, the influence of the sawtooth effect on depolarization
is small for the Z-mode and W-mode, but can be more dramatic for the H-mode.
In practice, the perturbation to the closed orbit and optics
as well as the spin resonance strengths
due to the sawtooth effect during the acceleration process
can be partly alleviated by ``tapering'' the magnetic field according to the local beam energy.  

In contrast with the Z-mode and W-mode, 
it is more challenging to maintain the
beam polarization in the acceleration to 120~GeV. 
For the booster lattice we studied in this paper,
the polarization loss due
to the intrinsic resonances could be reduced 
with a dedicated correction of the vertical equilibrium
emittance, while reducing the depolarization
due to the imperfection resonances requires
a better control of 
misalignment errors of quadrupoles,
as well as improvements in the strategy of closed-orbit correction.
Alternatively, the sensitivity
of the lattice structure to misalignment
errors in terms of the strengths of imperfection resonances
should be studied, as an important
ingredient in the lattice design and optimization. 
Nevertheless, even
if the depolarization effects of the imperfection
and intrinsic resonances can be well controlled,
alternative depolarization sources,
for example the synchrotron oscillations and radiative depolarization effects, 
could still cause depolarization near the extraction energy when
the acceleration rate becomes lower. These effects as well as the optimization of the
energy ramping curve requires further investigations.

Maintaining beam polarization in the CEPC booster
without the need for additional hardware,
such as Siberian snakes, would immediately make
injecting highly polarized beam(s)
into the collider rings 
an attractive solution for RD measurements and longitudinally polarized
colliding beam experiments. This approach has
the potential of achieving a higher level of beam
polarization, for both colliding and non-colliding pilot bunches,
compared to using
the Sokolov-Ternov effect in the collider rings.
This work endorses a careful study of different 
aspects of this approach, specifically for Z and W energies.

\section*{Acknowledgments}
The authors are grateful to D. P. Barber for his
helpful suggestions and careful reading of this manuscript, to S. Nikitin for 
discussions on experiments of polarized electron beam acceleration,
to D. Sagan and E. Forest for kind
help with Bmad/PTC. This study
was supported by
National Key Program for S\&T Research and Development (Grant No. 2018YFA0404300);
National Natural Science Foundation of China (Grant No. 11975252 and 12275283); Youth Innovation Promotion Association CAS (No. 2021012).


\bibliography{apssamp}

\end{document}